\documentclass{article}
\usepackage{graphicx}
\usepackage[centertags]{amsmath}
\usepackage{amsfonts}
\usepackage{amssymb}
\usepackage{amsthm}

\title{Quantized Mechanics of Affinely-Rigid Bodies}

\author{J.~J.~S\l awianowski, V.~Kovalchuk,\\
B.~Go\l ubowska, A.~Martens, E.~E.~Ro\.{z}ko\\
Institute of Fundamental Technological Research,\\
Polish Academy of Sciences,\\
$5^{\rm B}$, Pawi\'{n}skiego str., 02-106 Warsaw, Poland\\
e-mail: jslawian@ippt.pan.pl, vkoval@ippt.pan.pl,\\ 
bgolub@ippt.pan.pl, amartens@ippt.pan.pl, erozko@ippt.pan.pl}

\begin{document}

\maketitle
\begin{abstract}
In this paper we develope the main ideas of the quantized version of affinely-rigid (homogeneously deformable) motion. We base our consideration on the usual Schr\"odinger formulation of quantum mechanics in the configuration manifold which is given, in our case, by the affine group or equivalently by the semi-direct product of the linear group ${\rm GL}(n,\mathbb{R})$ and the space of translations $\mathbb{R}^{n}$, where $n$ equals the dimension of the ``physical space''. In particular, we discuss the problem of dynamical invariance of the kinetic energy under the action of the whole affine group, not only under the isometry subgroup. Technically, the treatment is based on the two-polar decomposition of the matrix of the internal configuration and on the Peter-Weyl theory of generalized Fourier series on Lie groups. One can hope that our results may be applied in quantum problems of nuclear dynamics or even in apparently exotic phenomena in vibrating neutron stars. And, of course, some more prosaic applications in macroscopic elasticity, structured continua, molecular dynamics, dynamics of inclusions, suspensions, and bubbles are also possible.
\end{abstract}

\section*{Introduction}

It is well known that every dynamical scheme, both mechanical and field-theoretical, may be formulated in two alternative versions: classical and quantum. This is always formally possible, although there are situations when one of those versions looks artificial, either classical or quantum. Besides, there are still some unsolved problems concerning the relationship between classical and quantum levels of description. Roughly speaking, they have to do with the measurement phenomena and decoherence processes. There is a whole spectrum of views, like, e.g.:
\begin{itemize}
\item[$(i)$] Everything is quantum and the classical is an illusion or approximation.
\item [$(ii)$] Everything should be finally classical and the quantum description is incomplete, phenomenological and temporary. Perhaps some essential nonlinearity may be a solution of paradoxes. 
\item [$(iii)$] Physical reality is a dualistic composition of two incompatible elements: quantum and classical. They are joined into a single whole in a way which is rather mysterious, although statistically described by the standard interpretation. This is a relatively popular view, in a sense philosophically based on a kind of the ``solliptic'' ideas.
\end{itemize}
But, as said above quite independently of those fundamental problems, almost every mechanical and field-theoretical scheme admits those two formulations: classical and quantum. For example, large molecules or fullerens may be described in a satisfactory way in both frameworks, every one to be used in its specific domain of applications. Moreover, such micro- and nano-objects are physically placed somewhere in the convolution region of two theories. What is then the relationship between them? This fact may perhaps shed some light onto the mentioned fundamental problems concerning the relationship and mutual interplay of both frameworks. It is not excluded that some basic questions concerning paradoxes and the relationship between classical and quantum levels of description might be solved or at least enlightened. For example, one can try to solve the problem by introducing to quantum considerations some geometrically motivated essential (non-perturbative) non-linearity \cite{DG-96,DGN-99,vk_03,vkjjs_08,jjs_12,jjsvk_02,jjsvk_10,jjsvk_13,all_12_1}. In any case, as many people seem to believe, this gives some hope for describing the decoherence phenomena. However, in this paper we are not intended to study the above fundamental problems. Our aim is mainly to formulate the quantum counterparts of the classical models studied earlier \cite{jjsvk_03,jjsvk_04,all_11,all_12_2,all_04}. 

\section{General scheme of the Schr\"odinger wave mechanics}

Let us begin with a short review of ideas of the Schr\"{o}dinger wave mechanics in a configuration manifold $Q$. 
Pure quantum states are described by wave functions on $Q$. We use the term ``wave functions'' although, as a matter of fact, they are not scalar functions but rather they are fields of complex scalar densities of weight $1/2$ on $Q$ \cite{Mac-63}. In principle they are assumed to be $L^{2}(Q)$-class fields on $Q$ in the sense of the natural scalar product
\begin{equation}\label{q.1}
( \Psi, \Phi )=\int \overline{\Psi}\Phi=\int \overline{\Psi}(q)\Phi(q)dq^{1} \ldots dq^{n},
\end{equation}
where $\Psi(q)$, $\Phi(q)$ are functions representing the densities $\Psi$, $\Phi$ in coordinates $q^{1}, \ldots , q^{n}$. We hope that our lack of rigour in distinguishing between global and local-coordinates expression for $(\Psi | \Phi)$ does not produce difficulty for mathemati\-cally-oriented readers. In any case, it is just intended for economy of notations. As said above, the rigorous formalism of quantum mechanics is based on the $L^{2}(Q)$-language. Nevertheless, in applications to the non-compact manifolds $Q$ one often does admit wave functions with the infinite $L^{2}(Q)$-norms, so-called scattering states. Obviously, then the quantity $P(\Psi, \Phi )=|(\Psi | \Phi)|^{2}$
does not possess the usual interpretation of the probability that the system in state $\Psi$ will be detected as one in the state $\Phi$. Nevertheless, the generalized statistical interpretation in terms of the relative probability is possible and physically convenient.

Physical quantities are represented by Hermitian operators in $L^{2}(Q)$. Their spectra are sets of their admissible values to be detected in the measurement processes. One can also admit non-bounded essentially self-adjoint operators. Moreover, such operators, often differential ones, are very important physically and represent such quantities like linear and angular momentum, position, energy etc. Obviously, their property to be self-adjoint is meant in the sense of the scalar product (\ref{q.1}), i.e., $(A \Psi | \Phi )=(\Psi | A \Phi)$
with certain provisos concerning the domain in the unbounded case. More generally, Hermitian conjugation is meant as $A \mapsto A^{+}$, where
$(A \Psi | \Phi )=(\Psi | A^{+} \Phi)$, again with provisos concerning domains when $A$ is non-bounded.

Unitary operators are symmetries of the scalar product (\ref{q.1}), 
$(U \Psi | U \Phi )=(\Psi | \Phi)$ for any wave functions $\Psi(q)$, $\Phi(q)$. To be more general, there are also important anti-unitary symmetries for which
$(U \Psi | U \Phi )=\overline{(\Psi | \Phi)}=(\Phi| \Psi)$. They are antilinear. For example, the quantum time reversal is anti-unitary. When taken together, two above-mentioned quantum automorphisms may be written down as
$|(U \Psi | U \Phi )|=|(\Psi | \Phi)|$. Unitary operators preserve the operation of Hermitian conjugation, i.e., $\left(U A U^{-1} \right)^{+}=U A^{+} U^{-1}$. The relationship between unitary and Hermitian operators is, roughly speaking, exponential:
\begin{equation}\label{q.9}
U=\exp (iA)=\sum_{n=0}^{\infty}\frac{1}{n!}(iA)^{n}, \qquad A^{+}=A.
\end{equation}
This is very important because some relationship is established between unbounded operators of physical quantities (like linear and angular momentum, position etc) and bounded unitary operators generated by them.

Eigenstates of the physical quantity $A$ satisfy the condition $A\Psi=a \Psi$,
where $a \in {\rm Sp} A \subset \mathbb{R} $ is an eigenvalue of $A$. In general the set of eigenvalues ${\rm Sp} A$ is a proper subset of the real axis. The above equation is to be solved both for $a$ and $\Psi$. Then (\ref{q.9}) implies that $U\Psi=\exp (ia) \Psi$, therefore, $\Psi$ is simultaneously an eigenstate of $U$. In practical problems one usually solves rather the first above eigenvalue problem than the second one.

This is a rough description of the Schr\"{o}dinger quantum scheme. However, in practical applications one uses a simplified description based on the scalar-valued wave functions. Indeed, when quantizing some classical theory, we usually begin with the classical expression for the kinetic energy:
\begin{equation}\label{q.12}
T=\frac{1}{2}\Gamma_{\mu\nu}\frac{dq^{\mu}}{dt}\frac{dq^{\nu}}{dt},
\end{equation}
where $q^{\mu}$ are local generalized coordinates in $Q$, and the metric coefficients $\Gamma_{\mu\nu}$ are functions of $q^{\mu}$. In general the metric $\Gamma$ is curved and there is no possibility to make $\Gamma_{\mu\nu}$ independent of $q^{\lambda}$ by any choice of coordinates. Let us also mention that the metric $\Gamma$ need not be positively definite. In pseudo-Riemannian manifold $(Q, \Gamma)$ there exists canonically distinguished measure $\mu_{\Gamma}$ induced by $\Gamma$ and given locally by:
\begin{equation}\label{q.13}
d\mu_{\Gamma}(q)=\sqrt{|{\rm det}[\Gamma_{\mu\nu}]|}dq^{1} \ldots dq^{n}.
\end{equation}
The factor $\sqrt{|{\rm det}[\Gamma_{\mu\nu}]|}$ is a scalar $W$-density of the weight $1$. Its square root is in consequence a scalar density of the weight $1/2$. Therefore, all scalar densities of this weight, in particular ``wave functions'' may be factorized as $\Psi(q)=\psi(q)\sqrt[4]{|{\rm det}\Gamma_{\mu\nu}|}$, where $\psi:Q \mapsto \mathbb{C}$ \ is a scalar-valued complex wave function on $Q$. The scalar product (\ref{q.1}) becomes $(\Psi|\Phi)=\langle\psi|\varphi\rangle=
\int\overline{\psi}(q)\varphi(q)d\mu_{\Gamma}(q)$ \cite{Mac-63}.
The action of operators on scalar $1/2$-densities $\Psi$ may be easily reformulated into the action on scalar wave functions $\psi$. And later on the whole Hilbert space formalism may be expressed in terms of the scalar product $\langle\psi| \varphi\rangle$. The classical kinetic energy (\ref{q.12}) may be easily expressed in Hamiltonian terms:
\begin{equation}\label{q.16}
\mathcal{T}=\frac{1}{2}\Gamma^{\mu\nu}p_{\mu}p_{\nu}, \qquad \Gamma^{\mu\alpha}\Gamma_{\alpha\nu}=\delta^{\mu}{}_{\nu}, \qquad p_{\mu}=\frac{\partial T}{\partial \dot{q}^{\mu}}=\Gamma_{\mu\nu}\frac{dq^{\nu}}{dt}.
\end{equation}
The corresponding quantum operator of kinetic energy and its Laplace-Beltrami operator are given by 
\begin{equation}\label{q.17-18}
\mathbf{T}=-\frac{\hbar^2}{2}\Delta(\Gamma),\qquad \Delta(\Gamma)=\frac{1}{\sqrt{|\Gamma|}}\sum_{\mu,\nu}
\partial_{\mu}\sqrt{|\Gamma|}\Gamma^{\mu\nu}\partial_{\nu}=\Gamma^{\mu\nu}
\nabla_{\mu}\nabla_{\nu}.
\end{equation}
In the last formula $\nabla_{\mu}$ is the Levi-Civita covariant differentiation in the sense of the metric tensor $\Gamma$. This means that the above quantum operator of kinetic energy is obtained from the classical expression (\ref{q.16}) by the mere substitution of the operator $\mathbf{P}_{\mu}=(\hbar/i) \nabla_{\mu}$ instead the classical momentum $p_{\mu}$. If some potential energy $V(q)$ is admitted, then the quantum Hamiltonian $\mathbf{H}$ is given by $\mathbf{H}=\mathbf{T}+\mathbf{V}$,
and $\mathbf{V}$ is the operator which multiplies wave functions by $V$,
$(\mathbf{V}\psi)(q)=V(q)\psi(q)$. Usually one does not distinguish graphically between the potential energy $V$ and its operator $\mathbf{V}$ given by the above formula. In a quite similar way one can include the magnetic interaction, using the minimal coupling procedure. It consists in introducing to the momentum operator $\mathbf{P}_{\mu}=(\hbar/i) \nabla_{\mu}$ some additive gauge correction linear in the covector potential of the magnetic field.

The non-relativistic quantum dynamics is based on Schr\"{o}dinger equation, i.e., $i\hbar \partial \psi/\partial t=\mathbf{H} \psi$. As usual, when $\mathbf{H}$ does not depend explicitly on time, it may be reduced to the time-independent Schr\"{o}dinger equation, i.e., to the eigenvalue problem for $\mathbf{H}$: $\mathbf{H} \varphi=E\varphi$, where both $E$ and $\varphi$ are a priori unknown, and in general the spectrum of the values of $E$ is not identical with the total real axis. Any solution of the above eigenvalue problem gives rise to the exponentially time-dependent solution of the Schr\"odinger equation, i.e., $\psi=\exp \left(-i E t/\hbar \right)\varphi$, where obviously $E$ is real, as an eigenvalue of the Hermitian operator $\mathbf{H}$. Therefore, the exponential time-dependence in $\psi$ is purely oscillatory.

Quantization conditions follow from the theory of Sturm-Liouville equations \cite{LL-58,Mes-65}. Namely, the admissible wave functions must be globally defined all over the configuration space, one-valued, and continuous together with their first-order derivatives. In the case of usual $L^{2}(Q, \mu)$-class functions describing bounded states, this demand implies the quantization condition for the admissible values of energy and other physical quantities. One obtains discrete spectra of eigenvalues of the corresponding operators. 

However, there are also some provisos and doubtful points here. The problem was noticed many years ago by W.~Pauli \cite{Pau-39}, J.~Reiss \cite{Rei-39} and also by D.~Arsenovi\'{c}, A.O.~Barut, M.~Bo\v{z}i\'{c}, Z.~Mari\v{c} \cite{ABB-95,ABMB-95,BBM-92}. Namely, there are situations when the configuration space $Q$ is multiply-connected and one can suspect that it is not wave function $\psi$ but rather $\overline{\psi}\psi$ what is to be one-valued. In any case this seems to be reasonable for such models as rigid body and affinely-rigid body in the physical space of dimension higher than $2$, because $Q$ is then doubly-connected and there are functions on the covering manifold $\overline{Q}$ which are non-projectable to $Q$, but nevertheless their squared moduli $\overline{\psi}\psi$ are correctly projectable. Namely, the values of $\psi$ differ in sign at points of $\overline{Q}$ projecting onto the same point of $Q$, therefore $\overline{\psi}\psi$ is a pull-back of some one-valued probability density on $Q$. Together with some superselection rule (no superposition of ``even'' and ``odd'' wave functions) this may lead to half-integer internal angular momentum, and therefore, to the classical ``explanation'' of spin. We shall return to this problem later on, but now we make some digression concerning the metric tensors and volume-measures on the configuration space $Q$.

\section{Specificity of Lie groups as configuration spa\-ces}

So, we assume that $Q$ is a Lie group or a Lie group-space \cite{Loo-63,Mac-63}. A similar treatment may be formulated for more general homogeneous spaces, i.e., such ones that the isotropy subgroups of any point are non-trivial, i.e., contain more elements than the identity alone. However, here we do not consider this more general treatment, because the configuration spaces of a rigid body or affinely-rigid body may be identified with Lie group manifolds, respectively ${\rm SO}(n,\mathbb{R})\widetilde{\times}\mathbb{R}^{n}$ and ${\rm GL}(n,\mathbb{R})\widetilde{\times}\mathbb{R}^{n}$. Let us remind that $n$ is the dimension of the physical space (i.e., $3$ but it is more convenient to assume it is arbitrary) and the sign $\widetilde{\times}$ denotes the semi-direct product.

The classical kinetic energy is given by (\ref{q.12}), (\ref{q.16}), and the quantum one is expressed by the Laplace-Beltrami operator (\ref{q.17-18}). This was a general manifold framework. However, when $Q$ is a Lie group, it is natural to assume that it is invariant under the left or right translations,
$x \mapsto gx$, $x \mapsto xg$, or perhaps under the both of them. In the case of metrically rigid body, $\mathcal{T}$ is spatially isotropic, i.e., invariant under the left translations $x \mapsto gx$. If the inertial tensor is isotropic, i.e., when we deal with the spherical rigid body, then $\mathcal{T}$ is invariant also under the right translations $x \mapsto xg$. Of course, this case of the material right-invariance is very special, but the left-invariance, i.e., spatial isotropy, is a general situation. So, the metric tensor $\Gamma$ on $Q$ is also left-invariant. But then also the measure $\mu_{\Gamma}$ (\ref{q.13}) is left-invariant. And now let us remind that on every locally-compact Lie group there exists left-invariant Haar measure and that it is unique up to normalization, i.e., up to a constant multiplicative factor. In the case of compact groups one can choose this factor in such a way that 
$\mu_{\Gamma}(Q)=\int_{Q}d \mu_{\Gamma}(q)=1$. 

All this is also true for the right-invariant Haar measure. And moreover, there is a wide class of groups, so-called unimodular groups for which the left-invariant Haar measure is identical with the right-invariant Haar measure. This class contains semi-simple groups, Abelian groups and their direct and semi-direct products. In any case the configuration spaces of a rigid and affine body belong here.

So, let us make a small digression concerning Hamiltonian systems on Lie groups and their quantization. Therefore, traditionally, we shall use the symbol $G$ instead $Q$, and the Lie algebra will be denoted by $G'$. Let $\Omega$, $\widehat{\Omega}\in G'$ be some elements of this algebra. They generate respectively some right-invariant and left-invariant vector fields $X[\Omega]$, $Y[\widehat{\Omega}]$. When the group $G$ is linear, they are given by 
$X[\Omega]_{g}=\Omega g$, $Y[\widehat{\Omega}]_{g}=g\widehat{\Omega}$.
In general, they are given by the tangent mappings of the right- and left-regular translations, $X[\Omega]_{g}=R_{g*}\Omega$, $Y[\widehat{\Omega}]_{g}=L_{g*}\widehat{\Omega}$, where $R_{g}(x)=xg$, $L_{g}(x)=gx$, where $\Omega$, $\widehat{\Omega}$ are arbitrary elements of the tangent space $T_{e}G$ at the group identity $e \in G$. 

In the case of linear groups, particularly interesting for us, the relationship between the Lie algebra $T_{e}G$ and other tangent spaces $T_{g}G$, e.g., $T_{g(t)}G$ where $t \mapsto g(t)$ is a curve in $G$, is given by 
$\Omega=(dg/dt)g(t)^{-1}$, $\widehat{\Omega}=g(t)^{-1}(dg/dt)=g(t)^{-1}\Omega g(t)$. In general we have $\Omega=R^{-1}_{g(t)*}(dg/dt)$, $\widehat{\Omega}=L_{g(t)*}^{-1}(dg/dt)$. Obviously, $R_{g*}$, $L_{g*}$ are tangent mappings of regular translations $R_{g}$, $L_{g}$. The momentum objects $\Sigma$, $\widehat{\Sigma}$ dual to $\Omega$, $\widehat{\Omega}$ are defined in the following way:
$\langle \Sigma, \Omega\rangle=\langle \widehat{\Sigma},\widehat{\Omega}\rangle =\langle p, \dot{g}\rangle$, where $\dot{g} \in T_{g}G$, $p \in T^{*}_{g}G$. The elements $g$, $\dot{g}$, $p$ in this definition are arbitrary. In absolute terms we can write $\Sigma=L_{g(t)}{}^{*}p$, $\widehat{\Sigma}=R_{g(t)}{}^{*}p$,
therefore $\Sigma=Ad_{g}{}^{*}\widehat{\Sigma}$. As usual, $L_{g}{}^{*}$, $R_{g}{}^{*}$, $Ad_{g}{}^{*}$ are pull-backs of $L_{g}{}_{*}$, $R_{g}{}_{*}$, $Ad_{g}{}_{*}$.

Transformation rules under regular translations $g \mapsto kg$, $g \mapsto gk$ read
\begin{eqnarray}
\Omega \mapsto Ad_{k}\Omega, \qquad \widehat{\Omega} \mapsto \widehat{\Omega},&\ & \Sigma \mapsto Ad_{k}{}^{*-1}\Sigma, \quad \: \widehat{\Sigma} \mapsto \widehat{\Sigma},\label{q.34a-c}\\
\Omega \mapsto \Omega, \quad\ \: \widehat{\Omega} \mapsto Ad^{-1}_{k}\widehat{\Omega},&\ & 
\Sigma \mapsto \Sigma, \qquad \widehat{\Sigma} \mapsto Ad_{k}{}^{*}\widehat{\Sigma}.\label{q.34b-d}
\end{eqnarray}
In the formulas above the symbol $Ad_{k}:G' \mapsto G'$ denotes the adjoint mapping, i.e., the derivative of the inner automorphism $g \mapsto kgk^{-1}$ at $g=e$, i.e., at the identity element. Let us complete those remarks by quoting the following obvious formulas for the right-invariant and left-invariant Maurer-Cartan forms $\eta$, $\omega$ on the linear group $G$: $\eta[\Sigma]_{g}=g^{-1}\Sigma$, $\omega[\widehat{\Sigma}]_{g}=\widehat{\Sigma}g^{-1}$. Let us remind also that for the potential model on a Lie group Hamiltonian is
\begin{equation}\label{q.36}
H=\mathcal{T}+\mathcal{V}(q)=\frac{1}{2}\Gamma^{\mu\nu}(q)p_{\mu}p_{\nu}+\mathcal{V}(q).
\end{equation}

Let $E_{\mu}$ denote some basic element in the Lie algebra $G'$ and $E^{\mu}$ are the corresponding elements of the dual basis in the space $G'^{*}$ of linear functions on $G'$. The canonical coordinates of the first kind are denoted by $q^{\mu}$, therefore, $g(q)=\exp (q^{\mu}E_{\mu})$.
The corresponding Lie-algebraic objects will be denoted by $\Omega$, $\widehat{\Omega} \in G'$ and their components by $\Omega^{\mu}$, $\widehat{\Omega}^{\mu}$, i.e., $\Omega=\Omega^{\mu}E_{\mu}$, $\widehat{\Omega}=\widehat{\Omega}^{\mu}E_{\mu}$. 
In the case of linear groups we have the obvious relationship with generalized velocities: $\Omega=(dg/dt)g^{-1}$, $\widehat{\Omega}=
g^{-1}(dg/dt)=g^{-1}\Omega g$, or in coordinates:
\begin{equation}\label{q.40}
\Omega^{\mu}=\Omega^{\mu}{}_{\nu}(q)\frac{dg^{\nu}}{dt}, \qquad \widehat{\Omega}^{\mu}=\widehat{\Omega}^{\mu}{}_{\nu}\frac{dg^{\nu}}{dt}.
\end{equation}
If $G$ is non-Abelian, then there are no coordinates $q^{\mu}$ in which $\Omega^{\mu}{}_{\nu}$ would be constant. The left- and right-invariant kinetic energies are given as follows:
\begin{equation}\label{q.41}
T_{\rm left}=\frac{1}{2}\mathcal{L}_{\mu\nu}(q)
\widehat{\Omega}^{\mu}\widehat{\Omega}^{\nu}, \qquad T_{\rm right}=\frac{1}{2}\mathcal{R}_{\mu\nu}(q)\Omega^{\mu}\Omega^{\nu}.
\end{equation}
The matrices $[\mathcal{L}_{\mu\nu}(q)]$, $[\mathcal{R}_{\mu\nu}(q)]$ here are constant, non-singular and symmetric. However, there is no need to assume them to be positively definite. 

For systems with the potential Hamiltonians (\ref{q.36}) one can describe Legendre transformation $\widehat{\Sigma}_{\mu}=\partial T_{\rm left}/\partial \widehat{\Omega}^{\mu}= \mathcal{L}_{\mu\nu}\widehat{\Omega}^{\nu}$,
$\Sigma_{\mu}=\partial T_{\rm right}/\partial\Omega^{\mu}=
\mathcal{R}_{\mu\nu}\Omega^{\nu}$ respectively for the left- and right-invariant kinetic energies. The quantities $\widehat{\Sigma}_{\mu}$, $\Sigma_{\mu}$ are components of the momentum mappings, i.e., Hamiltonian generators $\widehat{\Sigma}$, $\Sigma$ of the groups of right and left regular translations: $\widehat{\Sigma}=\widehat{\Sigma}_{\mu}E^{\mu}$, $\Sigma=\Sigma_{\mu}E^{\mu}$. Let us denote by $[\mathcal{L}^{\mu\nu}]$, $[\mathcal{R}^{\mu\nu}]$ the matrices reciprocal to $[\mathcal{L}_{\mu\nu}]$, $[\mathcal{R}_{\mu\nu}]$. One can show that the Hamiltonian (\ref{q.36}) may be written as follows:
\begin{eqnarray}
H_{\rm left} &= & \mathcal{T}_{\rm left}+\mathcal{V}\left(q\right)=\frac{1}{2}
\mathcal{L}^{\mu\nu}\widehat{\Sigma}_{\mu}\widehat{\Sigma}_{\nu}+
\mathcal{V}\left(q\right),\label{q.44a}\\
H_{\rm right}&= & \mathcal{T}_{\rm right}+\mathcal{V}\left(q\right)=\frac{1}{2}\mathcal{R}^{\mu\nu}
\Sigma_{\mu}\Sigma_{\nu}+\mathcal{V}\left(q\right)\label{q.44b}
\end{eqnarray}
respectively for the left- and right-invariant kinetic energy. These
expressions are formally quite analogous to the known expressions
in terms of canonical momenta. Nevertheless, their structure is completely
different from them if $G$ is not Abelian. Namely, $\widehat{\Omega}^{\mu}$,
$\Omega^{\mu}$ are non-holonomic quasi-velocities and $\widehat{\Sigma}_{\mu},\,\Sigma_{\nu}$
are quasimomenta. Their Poisson brackets do not vanish, namely, they
are expressed in the following way: 
\begin{equation}
\left\{ \Sigma_{\mu},\Sigma_{\nu}\right\} =\Sigma_{\lambda}C^{\lambda}\!_{\mu\nu},\qquad \left\{ \widehat{\Sigma}_{\mu},\widehat{\Sigma}_{\nu}\right\} =-\widehat{\Sigma}_{\lambda}C^{\lambda}\!_{\mu\nu},\qquad \left\{ \Sigma_{\mu},\widehat{\Sigma}_{\nu}\right\} =0,\label{q.45}
\end{equation}
where $C^{\lambda}\!_{\mu\nu}$ are structure constants of $G$ in the basis
composed of  $E_{\mu}$, i.e., $\left[E_{\mu},E_{\nu}\right]=E_{\lambda}C^{\lambda}\!_{\mu\nu}$.
Let us now describe the quantum counterparts of those objects and
concepts. First, let us remind that in majority of applications the
configuration space is a Lie group $Q=G$ or some of its homogeneous
spaces. For us it is sufficient to consider systems on Lie groups
or on their group manifolds. By a group manifold we mean its homogeneous
space with the trivial isotropy group $\left\{ e\right\} $ of every
point. Roughly speaking, it is a ``group which forgot about its identity
element''. We assume that $G$ is unimodular, i.e., the left-invariant
and right-invariant Haar measures coincide, thus there is only one
Haar measure $\mu$. It is possible to live without this assumption,
however in our models it is satisfied, so we accept it. The Hilbert
space of wave functions is $L^{2}(G,\mu)$. The left and right regular
translations in G generate in a natural way the following translation
operators in $L^{2}(G,\mu)$:
\begin{equation}
\left(\mathbf{L}(k)\Psi\right)(g)=\Psi(kg),\qquad
\left(\mathbf{R}(k)\Psi\right)(g)=\Psi(gk).\label{q.47}
\end{equation}
Translational invariance of the Haar measure implies that those operators
are unitary, therefore also bounded:
\begin{equation}
\left\langle \mathbf{L}(k)\Psi|\mathbf{L}(k)\varphi\right\rangle =\left\langle \Psi|\varphi\right\rangle =\left\langle \mathbf{R}(k)\Psi|\mathbf{R}(k)\varphi\right\rangle.\label{q.48}
\end{equation}
If the usual, or rather commonly used, definition of superposition
is used, $\left(F\circ G\right)(x)=F\left(G(x)\right)$, then $k\rightarrow\mathbf{R}(k)$, $k\rightarrow\mathbf{L}(k)$ are
respectively the representation and anti-representation of $G$ in
our Hilbert space $L^{2}(G,\mu)$: $\mathbf{R}(kl)=\mathbf{R}(k)\mathbf{R}(l)$,
$\mathbf{L}(kl)=\mathbf{L}(l)\mathbf{L}(k)$. It is clear that $\mathbf{L}$ becomes a representation when $k$ in (\ref{q.47}) is replaced by $k^{-1}$, i.e., if $\Psi\left(kg\right)$ is replaced by $\Psi\left(k^{-1}g\right)$. 
There are some delicate points concerning the infinitesimal representation
of unitary operators $\mathbf{L}(k)$, $\mathbf{R}(k)$. Namely, let
us define the following differential operators acting on differentiable
functions on $G$:
\begin{equation}
\left(\mathbf{L}_{\mu}f\right)(g)=  \left.\frac{\partial}{\partial q^{\mu}}f\left(k(q)g\right)\right|_{q=0},\qquad
\left(\mathbf{R}_{\mu}f\right)(g)=  \left.\frac{\partial}{\partial q^{\mu}}f\left(gk(q)\right)\right|_{q=0},\label{q.51a-b}
\end{equation}
where $q^{\mu}$ are canonical coordinates of the first kind, i.e.,
$g(q)=\exp (q^{\mu}E_{\mu})$.

The above-mentioned anti-representation and representation properties
of the groups of regular translations imply that 
$\left[\mathbf{L}_{\mu},\mathbf{L}_{\nu}\right]=
-\mathbf{L}_{\varkappa}C^{\varkappa}\!_{\mu\nu}$,
$\left[\mathbf{R}_{\mu},\mathbf{R}_{\nu}\right]=
\mathbf{R}_{\varkappa}C^{\varkappa}\!_{\mu\nu}$. The left and right regular translations mutually commute, therefore, $\left[\mathbf{L}_{\mu},\mathbf{R}_{\nu}\right]=0$. It is also clear that the Poisson brackets between $\Sigma_{\mu}$, $\widehat{\Sigma}_{\mu}$
and the functions $f$ which are pull-backs from the configuration
space (so that they depend only on the configuration variables $q^{\mu}$
but are independent of canonical momenta) may be expressed through the operators $\mathbf{L}_{\mu}$, $\mathbf{R}_{\mu}$:
\begin{equation}
\left\{ \Sigma_{\mu},f\right\} =-\mathbf{L}_{\mu}f,\qquad
\left\{ \widehat{\Sigma}_{\mu},f\right\}=-\mathbf{R}_{\mu}f.\label{q.54}
\end{equation}
Any other Poisson bracket may be expressed through the above ones
as 
\begin{equation}
\left\{ A,B\right\} =\Sigma_{\lambda}C^{\lambda}\!_{\mu\nu}\frac{\partial A}{\partial\Sigma_{\mu}}\frac{\partial B}{\partial\Sigma_{\nu}}-\frac{\partial A}{\partial\Sigma_{\mu}}\mathbf{L}_{\mu}B+\frac{\partial B}{\partial\Sigma_{\mu}}\mathbf{L}_{\mu}A,\label{q.55}
\end{equation}
when $A,B$ are treated as functions of $\Sigma_{\mu},q^{\nu}$. And
if we express them as functions of $\widehat{\Sigma}_{\mu}$, $q^{\nu}$,
then the following form is obtained:
\begin{equation}
\left\{ A,B\right\} =-\widehat{\Sigma}_{\lambda}C^{\lambda}\!_{\mu\nu}\frac{\partial A}{\partial\widehat{\Sigma}_{\mu}}\frac{\partial B}{\partial\widehat{\Sigma}_{\nu}}-\frac{\partial A}{\partial\widehat{\Sigma}_{\mu}}\mathbf{R}_{\mu}B+\frac{\partial B}{\partial\widehat{\Sigma}_{\mu}}\mathbf{R}_{\mu}A.\label{q.56}
\end{equation}
It is clear that the finite action of unitary operators may be expressed
in exponential way: $F\left(k(q)g\right)=\exp\left(q^{\mu}\mathbf{L}_{\mu}\right)F$, $F\left(gk(q)\right)=\exp\left(q^{\mu}\mathbf{R}_{\mu}\right)F$.
In this formula it is assumed that $F$ is smooth, i.e., $C^{\infty}$-class,
and the series on the right-hand side are convergent. Unlike this,
the left-hand side of the above formulas are always well defined, for any, even
drastically discontinuous functions $F$ on $G$. Obviously , in this
case the above formulas become false, because the right-hand sides
do not exist in the literal sense.

Let us quote the explicit formula for the differential operators $\mathbf{L}_{\mu}$, $\mathbf{R}_{\mu}$. The classical formulas (\ref{q.40}) for quantities imply that $\Sigma_{\mu}=p_{\alpha}\Sigma^{\alpha}\!_{\mu}$,
$\widehat{\Sigma}_{\mu}=p_{\alpha}\widehat{\Sigma}^{\alpha}\!_{\mu}$, where the matrices $[\Sigma^{\alpha}\!_{\mu}]$, $[\widehat{\Sigma}^{\alpha}\!_{\mu}]$
are reciprocal to $[\Omega^{\mu}\!_{\alpha}]$, $[\widehat{\Omega}^{\mu}\!_{\alpha}]$, i.e.,
$\Sigma^{\alpha}\!_{\mu}\Omega^{\mu}\!_{\beta}=\delta^{\alpha}\!_{\beta}$,
$\widehat{\Sigma}^{\alpha}\!_{\mu}\widehat{\Omega}^{\mu}\!_{\beta}=
\delta^{\alpha}\!_{\beta}$. Therefore, our differential operators $\mathbf{L}_{\mu}$, $\mathbf{R}_{\mu}$
are given by $\mathbf{L}_{\mu}=\Sigma^{\alpha}\!_{\mu}\partial/\partial q^{\alpha}$, $\mathbf{R}_{\mu}=\widehat{\Sigma}^{\alpha}\!_{\mu} \partial/\partial q^{\alpha}$. The formulas (\ref{q.48}) tell us that $\mathbf{L}(k)$, $\mathbf{R}(k)$ are unitary with respect to the scalar product based on the Haar measure $\mu$ on $G$. Therefore we have that
$\mathbf{L}\left(\exp\left(q^{\mu}E_{\mu}\right)\right)=
\exp\left(q^{\mu}\mathbf{L}_{\mu}\right)$,
$\mathbf{R}\left(\exp\left(q^{\mu}E_{\mu}\right)\right)=
\exp\left(q^{\mu}\mathbf{R}_{\mu}\right)$, and that $\mathbf{L}_{\mu}$, $\mathbf{R}_{\mu}$ are formally anti-self-adjoint:
\begin{equation}
\left\langle \mathbf{L}_{\mu}\Psi|\varphi\right\rangle =-\left\langle \Psi|\mathbf{L}_{\mu}\varphi\right\rangle,\qquad \left\langle \mathbf{R}_{\mu}\Psi|\varphi\right\rangle =-\left\langle \Psi|\mathbf{R}_{\mu}\varphi\right\rangle.\label{q.62}
\end{equation}
 
As, mentioned above, due to the fact that $\mathbf{L}_{\mu}$, $\mathbf{R}_{\mu}$
are differential operators, one must be careful with (\ref{q.62}). Namely,
only for $\Psi$, $\varphi$ from the appropriate domain this equation
holds. In any case, it is certainly true for infinitely smooth functions
with compact supports, i.e., for $\Psi,\varphi\in C_{0}^{\infty}(G)$.
Let us mention that in spite of the algebraic rule (\ref{q.62}), the operators
$\mathbf{L}_{\mu}$, $\mathbf{R}_{\mu}$ are not anti-hermitian because
they are non-bounded and defined only on some domain of $L^{2}(G,\mu)$.
Nevertheless, in the case of operators like $\mathbf{L}_{\mu}$, $\mathbf{R}_{\mu}$ which are infinitesimal generators of well-defined unitary groups, the danger of misunderstandings following from interpreting them as
``anti-hermitian'' is much smaller than that for quite arbitrary, so-to-speak accidental differential operators. The operators 
$\mathbf{\Sigma}_{\mu}:=(\hbar/i)\mathbf{L}_{\mu}$,
$\widehat{\mathbf{\Sigma}}_{\mu}:=(\hbar/i)\mathbf{R}_{\mu}$
are counterparts of $\Sigma_{\mu}$, $\widehat{\Sigma}_{\mu}$. They
are ``hermitian'' with identical provisos as $\mathbf{L}_{\mu}$, $\mathbf{R}_{\mu}$ are ``anti-hermitian'', thus, 
$\left\langle \mathbf{\Sigma}_{\mu}\Psi|\varphi\right\rangle =\left\langle \Psi|\mathbf{\Sigma}_{\mu}\,\varphi\right\rangle$, $\langle \widehat{\mathbf{\Sigma}}_{\mu}\Psi|\varphi\rangle =\langle \Psi|\widehat{\mathbf{\Sigma}}_{\mu}\varphi\rangle$ for any $\Psi,\varphi\in C_{0}^{\infty}(G)$. Obviously, the operators $\mathbf{\Sigma}_{\mu}$, $\widehat{\mathbf{\Sigma}}_{\mu}$ may be expressed as
\begin{equation}
\mathbf{\Sigma}_{\mu}:=\frac{\hbar}{ i }\Sigma^{\alpha}\!_{\mu}(q)\frac{\partial}{\partial q^{\alpha}},\qquad \widehat{\mathbf{\Sigma}}_{\mu}:=\frac{\hbar}{ i }\widehat{\Sigma}^{\alpha}\!_{\mu}(q)\frac{\partial}{\partial q^{\alpha}}.\label{q.65}
\end{equation}

For any pair of physical quantities represented on the quantum level
by ``hermitian'' (essentially self-adjoint) operators $\mathbf{F}$, 
$\mathbf{G}$ the quantum Poisson bracket is defined as 
${}_{Q}\left\{ \mathbf{F},\mathbf{G}\right\} =(1/i\hbar)\left[\mathbf{F},\mathbf{G}\right]=
(1/i\hbar)\left(\mathbf{F}\mathbf{G}-\mathbf{G}\mathbf{F}\right)$,
i.e., as the commutator divided by $i\hbar$. Therefore, we see that the quantum Poisson brackets for $\mathbf{\Sigma}_{\mu},\;\widehat{\mathbf{\Sigma}}_{\mu}$
have the form analogous with the classical one (\ref{q.45}): 
\begin{equation}
_{Q}\left\{ \mathbf{\Sigma}_{\mu},\mathbf{\Sigma}_{\nu}\right\} =\mathbf{\Sigma}_{\lambda}C^{\lambda}\!_{\mu\nu},\quad {}_{Q}\left\{ \widehat{\mathbf{\Sigma}}_{\mu},\widehat{\mathbf{\Sigma}}_{\nu}\right\} =-\mathbf{\Sigma}_{\lambda}C^{\lambda}\!_{\mu\nu},\quad {}_{Q}\left\{ \mathbf{\Sigma}_{\mu},\widehat{\mathbf{\Sigma}}_{\nu}\right\} =0.\label{q.67}
\end{equation}

The quantum counterparts of (\ref{q.41}), i.e., operators of the left-invariant
and right invariant kinetic energy are 
$\mathbf{T}_{\rm left}=(1/2)\mathcal{L}^{\mu\nu}\widehat{\mathbf{\Sigma}}_{\mu}
\widehat{\mathbf{\Sigma}}_{\nu}=-(\hbar^{2}/2)
\mathcal{L}^{\mu\nu}\mathbf{R}_{\mu}\mathbf{R}_{\nu}$,
$\mathbf{T}_{\rm right}=(1/2)\mathcal{R}^{\mu\nu}\mathbf{\Sigma}_{\mu}
\mathbf{\Sigma}_{\nu}=-(\hbar^{2}/2)\mathcal{R}^{\mu\nu}
\mathbf{L}_{\mu}\mathbf{L}_{\nu}$, respectively. Let us repeat that in rigid body mechanics (let us stress: metrically rigid) one is dealing with $\mathbf{T}_{\rm left}$, i.e., with the kinetic energy invariant under spatial rotations. In general, it
is not right-, i.e., metrically-invariant. More precisely, it is right-invariant
only when $\mathcal{L}_{\mu\nu}\left(\mathcal{L}^{\mu\nu}\right)$
is so, i.e., if it is proportional to the unit tensor, $\mathcal{L}^{\mu\nu}=(1/I)\delta^{\mu\nu}$. An intermediary symmetry in $n=3$ occurs when the inertial tensor $\mathcal{L}^{\mu\nu}$ is once degenerate with respect to $\delta^{\mu\nu}$, i.e., when two mean values (mean moments of inertia) coincide. Let us remind those mean values are solutions of the eigenequation: $\det\left[\mathcal{L}^{\mu\nu}-\lambda\delta^{\mu\nu}\right]=0$.

Unlike the rigid body mechanics, the Hamiltonian model of the ideal
incompressible fluid is right-invariant. Let us mention another fundamental
difference between rigid body and incompressible fluid. Rigid body
without translational motion has $n(n-1)/2$ degrees of freedom,
and with translational motion $n(n+1)/2$, so for $n=3$ respectively
3 and 6. In any case it is a finite number. Incompressible ideal fluid
has the infinite-continuous number of degrees of freedom. Namely, the configuration space is given by the infinite-dimensional manifold of all volume-preserving diffeomorphisms. And the kinetic energy is invariant on right under the group of all such diffeomorphisms. On the left it is invariant
only under the group of Euclidean isometries, obviously finite-dimensional
one $n(n+1)/2$. Incidentally, let us mention that the theory
of infinite-dimensional ``Lie groups'' is far from being complete,
and in any case it is rather difficult. Nevertheless the structure
of Hamiltonian or quantum mechanics on it is a good, very convenient tool for guessing some solutions; once guessed in this way, they may be proved by independent methods. Obviously, the ``full happiness'' appears when the kinetic energy is doubly-invariant, i.e., both left and right. Obviously, it is the case for Abelian groups, but more interesting is the case of semi-simple groups. Then one can use the Killing tensor on $G$ as the metric tensor. It is so, e.g., for the spherical rigid body. Let us remind the definition of the Killing tensor on the Lie algebra $G'$, i.e., $\gamma_{\mu\nu}=C^{\alpha}{}_{\beta\mu}C^{\beta}{}_{\alpha\nu}$, 
and its extension to the pseudo-Riemannian structure on the manifold $G$:
$\Gamma_{\mu\nu}(q)=\gamma_{\alpha\beta}\Sigma^{\alpha}{}_{\mu}(q)
\Sigma^{\beta}{}_{\nu}(q) =\gamma_{\alpha\beta}\widehat{\Sigma}^{\alpha}{}_{\mu}(q)
\widehat{\Sigma}^{\beta}{}_{\nu}(q)$. To be more precise: if $G$ is the Cartesian product of a few simple groups, $G=G_{1} \times \ldots \times G_{p} = \times_{k=1}^{N}G_{k}$, then one can use the Killing tensors $\Gamma(k)$ on each $G_{k}$, and combine them with constant coefficients: $\Gamma=\sum^{N}_{k=1}C_{k}\pi_{k}{}^{\ast}\Gamma(k)= 
C_{1}\pi_{1}{}^{\ast}\Gamma(1)+ 
\ldots +C_{N}\pi_{N}{}^{\ast}\Gamma(N)$, where $\pi_{k}=G \rightarrow G_{k}$ 
denotes the natural projections onto factors of the Cartesian product. A similar procedure may be used when $G$ is a direct or semi-direct product of the semisimple group with another, usually Abelian group with some physically motivated metric structure.

Let us complete (\ref{q.67}) by the quantum counterpart of (\ref{q.54}). Namely, let $F$ be a function on the configuration space. It gives rise to the following multiplication operator on wave functions: $({\mathbf F}\Psi)(q)=F(q)\Psi(q)$. 
Then one can easily show that the following quantum Poisson-bracket rules are satisfied: ${}_{Q}\left\{{\boldsymbol\Sigma}_{\mu}, {\mathbf F}\right\}=-{\mathbf L}_{\mu}F$, ${}_{Q}\{\widehat{{\boldsymbol\Sigma}}_{\mu}, {\mathbf F}\}=-{\mathbf R}_{\mu}F$. Just as (\ref{q.67}) they are formally analogous to the classical rules. However, one must remember that it is an exceptional situation which holds only for special quantities of geometric origin and with clear and simple geometric interpretation. In general such a quantum-classical correspondence does not exist. Those were some relatively general remarks. Now we go back to the mechanics of affinely-rigid body.

\section{Quantization of affinely-rigid bodies}

 As it was said a few times above, the configuration space of an affinely-rigid body may be identified with ${\rm LI}(U,V)\times M$, the Cartesian product of the internal configuration space and the manifold of translational degrees of freedom
\cite{vk_10,eer_10,jjs_07,jjsvk_04,all_10,all_05}. When we choose some orthonormal Cartesian coordinates in $M$, $N$, namely $x^{i}$, $a^{K}$, then the induced coordinates in the configuration space are $x^{i}$, $\varphi^{i}{}_{K}$ and the configuration manifold itself is identified with the semi-direct product ${\rm GL}(n,{\mathbb R})\widetilde{\times}{\mathbb R}^{n}$. Apparently the most natural measures on ${\rm GL}(n,{\mathbb R})\widetilde{\times}{\mathbb R}^{n}$ and ${\rm GL}(n,{\mathbb R})$ are $a$, $l$, where $da(\varphi, x)=dx^{1}\ldots dx^{n}d\varphi^{1}{}_{1}\ldots d\varphi^{n}{}_{n}=dl(\varphi)dx^{1}\ldots dx^{n}$,
$dl(\varphi)=d\varphi^{1}{}_{1}\ldots d\varphi^{n}{}_{n}$. Obviously, they are not Haar measures in the group-theoretic sense, although they are Lebesgue measures in the additive sense of ${\rm L}(n,{\mathbb R})\times {\mathbb R}^{n}$. The corresponding Haar measures $\alpha$, $\lambda$ are given as follows: $d\alpha(\varphi, x)=(\det\varphi)^{-n-1}da(\varphi,x)$,
$d\lambda(\varphi)=(\det\varphi)^{-n}dl(\varphi)$. In practical problems, first of all when dealing with non-geodetic models, i.e., ones containing the potential term in Lagrangian, it is convenient to express those measures in terms of the two-polar splitting variables. Let $\nu$ denote the Haar measure on the manifold of orthonormal frames, or equivalently --- on the rotation group ${\rm SO}(n,{\mathbb R})$ \cite{Wey-28,Wey-31,Wig-31,Wig-65}. Then the above formula takes on the following intuitive and suggestive form in terms of the two-polar decomposition:
\begin{equation}
d\lambda(\varphi)=d\lambda(L,q,R)=\prod_{i\neq j}\left|{\rm sh}\left(q^{i}-q^{j}\right)\right|d\nu(L)d\nu(R)dq^{1}\ldots dq^{n}. \label{q.80}
\end{equation}
Let us remind that the two-polar splitting is meant in the sense
$\varphi=LDR^{-1}$, $D= {\rm Diag}\left(\ldots , Q^{a},\ldots\right)={\rm Diag}\left(\ldots , \exp q^{a},\ldots\right)$. In models invariant only under the Euclidean group, it is more convenient to use the $Q$-variables. Then  $dl(\varphi)=dl(L,Q,R)=\prod_{i\neq j}\left|\left(Q^{i}+Q^{j}\right)\left(Q^{i}-Q^{j}\right)\right|
d\nu(L)d\nu(R)dQ^{1}\ldots dQ^{n}$. It is often convenient to introduce the shortened symbols:
\begin{equation}
P_{\lambda}=\prod_{i\neq j}\left|{\rm sh}\left(q^{i}-q^{j}\right)\right|, 
\qquad P_{l}=\prod_{i\neq j}\left|\left(Q^{i}+Q^{j}\right)\left(Q^{i}-Q^{j}\right)\right|. \label{q.83}
\end{equation}
Then the above formulas read as follows:
$d\lambda(\varphi)=P_{\lambda}d\nu(L)d\nu(R)dq^{1}\ldots dq^{n}$, $dl(\varphi)=P_{l}d\nu(L)d\nu(R)dQ^{1}\ldots dQ^{n}$. 

Let us observe that in the analysis of affinely-invariant systems we often separate the motion into the ${\rm SL}(n,{\mathbb R})$ and purely dilatational part. Then it is convenient to use the Haar measure on the ${\rm SL}(n,\mathbb{R})$-part of motion. It may be symbolically expressed in terms of the ``delta-distribution''-based expression:
$d\lambda_{{\rm SL}}(\varphi)=P_{\lambda}d\nu(L)d\nu(R)\delta(q^{1}+\ldots +q^{n})dq^{1}\ldots dq^{n}$. The factor $\delta (q^{1}+\ldots +q^{n})$ switches out the integration along the dilatational parameter and reduces the procedure to the submanifold of isochoric motions. 

Let us now translate our general group formulas just to the above language of affinely-rigid body, i.e., to the situation when the group ${\rm G}$ is given by ${\rm GL}(n,{\mathbb R})$. First of all, our index $\mu$ becomes the two-index $({}^{a}{}_{b})$. The affine spin in laboratory and co-moving representations are given by the following differential operators:
\begin{equation}
{\boldsymbol\Sigma}^{a}{}_{b}:=\frac{\hbar}{i}{\mathbf L}^{a}{}_{b}=\frac{\hbar}{i}\varphi^{a}{}_{K}\frac{\partial}{\partial \varphi^{b}{}_{K}}, \qquad  \widehat{{\boldsymbol \Sigma}}^{A}{}_{B}:=\frac{\hbar}{i}{\mathbf R}^{A}{}_{B}=\frac{\hbar}{i}\varphi^{m}{}_{B}\frac{\partial}{\partial \varphi^{m}{}_{A}}. \label{q.86}
\end{equation}
And the metrical spin and vorticity are represented by their $g$- and $\eta$-skew-symmetric parts:
${\mathbf S}^{a}{}_{b}={\boldsymbol\Sigma}^{a}{}_{b}-g^{ac}g_{bd}
{\boldsymbol\Sigma}^{d}{}_{c}$, ${\mathbf V}{}^{A}{}_{B}= \widehat{{\boldsymbol \Sigma}}{}^{A}{}_{B}-\eta^{AC}\eta_{BD}\widehat{{\boldsymbol \Sigma}}{}^{D}{}_{C}$. 

\noindent{\bf Remark:} we were just using two measures $\lambda$, $l$ on the manifold of internal degrees of freedom. Only $\lambda$ is the Haar measure, but as mentioned, $l$ and the corresponding Hilbert space ${\rm L}^{2}(Q,l)$ may be also useful, namely in non-affine models of internal symmetries. But then, in the Hilbert space, (\ref{q.86}) fail to be formally self-adjoint. Instead, we would have to modify them by introducing some additive corrections, i.e.,
${\boldsymbol\Sigma}(l)^{a}{}_{b}={\boldsymbol\Sigma}^{a}{}_{b}+(\hbar n/2i)\delta^{a}{}_{b}$, $\widehat{{\boldsymbol \Sigma}}(l)^{A}{}_{B}=\widehat{{\boldsymbol \Sigma}}{}^{A}{}_{B}+(\hbar n/2i)\delta^{A}{}_{B}$. It is clear that this correction influences only the dilatational part of generators. In particular, the spin and vorticity operators do not feel anything. Obviously, we have the following well-known classical identity: ${\boldsymbol\Sigma}^{a}{}_{b}=\varphi^{a}{}_{A}\varphi^{-1B}{}_{b}
\widehat{{\boldsymbol \Sigma}}{}^{A}{}_{B}$. Nevertheless, just like in the classical model, ${\mathbf V}^{A}{}_{B}$ are not co-moving components of spin ${\mathbf S}^{a}{}_{b}$.

The operators of translational linear momentum, respectively in the spatial and co-moving representation, ${\mathbf P}_{a}$ and $\widehat{\mathbf P}_{A}$ are given by ${\mathbf P}_{a}=(\hbar/i)\partial/\partial x^{a}$, $\widehat{{\mathbf P}}_{A}=(\hbar/i)\varphi^{a}{}_{A}\partial/\partial x^{a}$. Obviously, they are also interrelated through $\varphi$, just like ${\boldsymbol\Sigma}^{a}{}_{b}$, $\widehat{{\boldsymbol \Sigma}}{}^{A}{}_{B}$, i.e., $\widehat{{\mathbf P}}_{A}=\varphi^{a}{}_{A}{\mathbf P}_{a}$, ${\mathbf P}_{a}=\varphi^{-1 A}{}_{a}\, \widehat{{\mathbf P}}_{A}$. Just like in the classical theory, the quantities ${\boldsymbol\Sigma}^{a}{}_{b}$, $\widehat{{\boldsymbol \Sigma}}{}^{A}{}_{B}$ are respectively infinitesimal generators of the left and right regular translations of $\varphi$: $\varphi^{a}{}_{A}\mapsto L^{a}{}_{b}\varphi^{b}{}_{A}$, $\varphi^{a}{}_{A}\mapsto \varphi^{a}{}_{B}R^{B}{}_{A}$. And similarly the Hermitian operators ${\mathbf P}_{a}$, $\widehat{{\mathbf P}}_{A}$ are generators of the spatial and material translations of the centre of mass position, e.g., $x^{a}\mapsto x^{a} + w^{a}$.
Exactly like in the classical case, we can also introduce the translational and total affine momenta with respect to some fixed origin ${\mathfrak O} \in M$,
i.e., ${\boldsymbol \Lambda}[{\mathfrak O}]^{i}{}_{j}=x^{i}{\mathbf P}_{j}$,  ${\mathbf J}[{\mathfrak O}]^{i}{}_{j}={\boldsymbol \Lambda}[{\mathfrak O}]^{i}{}_{j} +{\boldsymbol \Sigma}^{i}{}_{j}$. They are Hermitian generators of the group ${\rm GAff}(M)$ acting, e.g., through $\left(x^{a},\varphi^{a}{}_{A}\right)\mapsto
\left(L^{a}{}_{b}x^{b},L^{a}{}_{b}\varphi^{b}{}_{A}\right)$, 
and similarly for the $x$-translations. The commutation rules are just like Poisson brackets in classical mechanics.

In analogy to the classical canonical momentum $p$ canonically conjugate to the centre of mass of logarithmic  deformation invariants, we introduce the following formally self-adjoint operator: ${\mathbf p}=(\hbar/i)\partial/\partial q$. It is a generator of the quantum group of dilatations. In certain formulas it is also convenient to use the shear (deviatoric) component of the affine spin, i.e.,
\begin{equation}
{\boldsymbol\sigma}^{a}{}_{b}:={\boldsymbol \Sigma}^{a}{}_{b}-\frac{1}{n}\,{\mathbf p}\,\delta^{a}{}_{b}, \qquad \widehat{{\boldsymbol \sigma}}^{A}{}_{B}:=\widehat{{\boldsymbol \Sigma}}^{A}{}_{B}-\frac{1}{n}\,{\mathbf p}\,\delta^{A}{}_{B}. \label{q.97}
\end{equation}
It is clear that the dilatational generator ${\mathbf p}$ equals the trace of the tensor operators ${\boldsymbol \Sigma}^{a}{}_{b}$, $\widehat{{\boldsymbol \Sigma}}{}^{A}{}_{B}$, i.e., ${\mathbf p}={\boldsymbol \Sigma}^{a}{}_{a}=\widehat{{\boldsymbol \Sigma}}{}^{A}{}_{A}$.

Just like in the classical theory, the spin and minus-vorticity operators are (formally) self-adjoint generators of the orthogonal subgroup ${\rm SO}(V,g)\subset {\rm GL}(V)$, ${\rm SO}(U,\eta)\subset {\rm GL}(U)$ acting on the configuration space of an affinely-rigid body. Equivalently, one can say that they generate left regular translations of the $L$ and $R$ factors in the two-polar decomposition, $\varphi=LDR^{-1}$. The Lie algebras of ${\rm SO}(V,g)$, ${\rm SO}(U,\eta)$ consist respectively of $g$- and $\eta$-skew-symmetric linear mappings. So, their matrix elements satisfy the rule:
\begin{equation}\label{q.99}
\mu^{i}{}_{j}=-g^{ik}g_{jl}\mu^{l}{}_{k},\qquad 
\nu^{A}{}_{B}=-\eta^{AC}\eta_{BD}\nu^{D}{}_{C}.
\end{equation}
The linear spaces of such mappings are spanned on the elements $E^{k}{}_{l}$, $H^{A}{}_{B}$:
\begin{equation}\label{q.100}
\left(E^{k}{}_{l}\right)^{i}{}_{j}=\delta^{k}{}_{j}\delta^{i}{}_{l}
-g^{ki}g_{lj},\qquad \left(H^{A}{}_{B}\right)^{C}{}_{D}=\delta^{A}{}_{D}\delta^{C}{}_{B}
-\eta^{AC}\eta_{BD}.
\end{equation}
They do not form basis because they are linearly dependent in a consequence of their antisymmetry properties (\ref{q.99}). But one can operate with them as if they were bases if expansion coefficients are subject to (\ref{q.99}) in all formulas. So, let $W(\mu)$, $T(\nu)$ be finite transformations from ${\rm SO}(V,g)$, ${\rm SO}(U,\eta)$, i.e.,
\begin{equation}\label{q.101}
W(\mu)=\exp\left(\frac{1}{2}\mu^{i}{}_{j}E^{j}{}_{i}\right),\qquad T(\nu)=\exp\left(\frac{1}{2}\nu^{A}{}_{B}H^{B}{}_{A}\right)
\end{equation}
with $\mu$, $\nu$ satisfying (\ref{q.99}). The factor 1/2 appears just because of this skew-symmetry assumption. For any functions $F$, $H$ of the $L,R$-arguments respectively the transformations (\ref{q.101}) act according to the following rules:
\begin{eqnarray}
F\left(W(\mu)L\right)&=&\left(\exp\left(\frac{i}{2\hbar}\mu^{i}{}_{j}
\mathbf{S}^{j}{}_{i}\right)F\right)(L), \label{q.102a}\\
H\left(T(\nu)R\right)&=&\left(\exp\left(-\frac{i}{2\hbar}\nu^{A}{}_{B}
\mathbf{V}^{B}{}_{A}\right)H\right)(R). \label{q.102b}
\end{eqnarray}
Therefore, roughly speaking, the spin and minus-vorticity operators correspond exactly to the classical quantities $\varrho$, $\tau$ which were Hamiltonian generators of spatial and material rigid rotations. Let us remind that in classical theory the $L$- and $R$-co-moving components $\widehat{\varrho}$, $\widehat{\tau}$ were Hamiltonian generators of ${\rm SO}(n,\mathbb{R})$ acting on the right on the $L,R$-factors. Similarly in quantized theory we introduce the $L$- and $R$-co-moving components of spin and vorticity:
${\boldsymbol\varrho}^{a}{}_{b}=L^{a}{}_{i}L^{j}{}_{b}\mathbf{S}^{i}{}_{j}$,
${\boldsymbol\tau}^{a}{}_{b}=-R^{B}{}_{b}R^{a}{}_{A}\mathbf{V}^{A}{}_{B}$.
Indices in the above equations are raised and lowered with the help of our metric tensors $g\in V^{\ast}\otimes V^{\ast} (i,j)$, $\eta\in U^{\ast}\otimes U^{\ast} (A,B)$ and the Kronecker delta $\delta\in \mathbb{R}^{n\ast}\otimes\mathbb{R}^{n\ast}(a,b)$. Let us remind also that in this convention $L^{j}{}_{b}$, $R^{A}{}_{b}$ are matrix elements of $L:\mathbb{R}^{n}\rightarrow V$, $R:\mathbb{R}^{n}\rightarrow U$, and $L^{a}{}_{i}$, $R^{a}{}_{B}$ are matrix elements of $L^{-1}:V\rightarrow\mathbb{R}^{n}$, $R^{-1}:U\rightarrow\mathbb{R}^{n}$.

The role of ${\boldsymbol\varrho}^{a}{}_{b}$, ${\boldsymbol\tau}^{a}{}_{b}$ as Hamiltonian generators of the right-hand-side translations of $L\in{\rm SO}\left(\mathbb{R}^{n},\delta;V,g\right)$, $R\in{\rm SO}\left(\mathbb{R}^{n},\delta;U,\eta\right)$ by elements of the orthogonal group ${\rm SO}(n,\mathbb{R})$ may be described by the formulas similar to (\ref{q.102a}), (\ref{q.102b}). Namely, we again use the system of matrices $\left(\epsilon^{a}{}_{b}\right)$ in the Lie algebra ${\rm SO}(n,\mathbb{R})^{\prime}$ analogous to (\ref{q.100}), i.e.,
$\left(\epsilon^{a}{}_{b}\right)^{i}{}_{j}=\delta^{a}{}_{j}\delta^{i}{}_{b}-
\delta^{ai}\delta_{bj}$, and represent the finite elements of ${\rm SO}(n,\mathbb{R})$ in the exponential form:
\begin{equation}\label{q.105}
Z(\omega)=\exp\left(\frac{1}{2}\omega^{a}{}_{b}\epsilon^{b}{}_{a}\right).
\end{equation}
The coefficients matrix $\left[\omega^{a}{}_{b}\right]$ is $\delta$-skew-symmetric, i.e., $\omega^{a}{}_{b}=-\delta^{ac}\delta_{bd}\omega^{d}{}_{c}$.
Then the action of (\ref{q.105}) on $L$, $R$ is represented in the linear space of functions of $L$ and $R$ variables as follows:
\begin{eqnarray}
F\left(LZ(\omega)\right)&=&\left(\exp\left(\frac{i}{2\hbar}\omega^{a}{}_{b}
{\boldsymbol\varrho}^{b}{}_{a}\right)F\right)(L), \label{q.107a}\\
H\left(RZ(\omega)\right)&=&\left(\exp\left(-\frac{i}{2\hbar}\omega^{a}{}_{b}
{\boldsymbol\tau}^{b}{}_{a}\right)H\right)(R). \label{q.107b}
\end{eqnarray}
One point must be stressed here. In formulas (\ref{q.102a}), (\ref{q.102b}), (\ref{q.107a}), (\ref{q.107b}) it is stated that the action of $W(\mu)$, $T(\nu)$ and the both actions of $Z(\omega)$ are expressed through the operator exponent acting on functions of $L$ or $R$ variables. And the exponentiated operators are linear in $\mathbf{S}^{i}{}_{j}$, $\mathbf{V}^{A}{}_{B}$, ${\boldsymbol\varrho}^{a}{}_{b}$, ${\boldsymbol\tau}^{a}{}_{b}$. But of course the two-polar expansion $\varphi=LDR^{-1}$ enables one to translate this immediately onto the exponent action on the total configuration space ${\rm AffI}(N,M)$. For infinitesimal group parameters $\mu^{i}{}_{j}\approx 0$, $\nu^{a}{}_{B}\approx 0$, $\omega^{a}{}_{b}\approx 0$, the increment of operators is again given by the quantum Poisson bracket with generator multiplied by the small group parameter. Therefore, up to the higher-order terms in parameters, i.e., in the linear approximation, the infinitesimal increments of operators under the action of transformation groups (\ref{q.102a}), (\ref{q.102b}), (\ref{q.107a}), (\ref{q.107b}) are given by
\begin{eqnarray}
\delta\mathbf{A}&=&\frac{1}{2}\mu^{i}{}_{j}\{\mathbf{A},\mathbf{S}^{j}{}_{i}
\}_{QPB}=\frac{1}{2i\hbar}\mu^{i}{}_{j}[\mathbf{A},\mathbf{S}^{j}{}_{i}], \label{q.108a}\\
\delta\mathbf{A}&=&\frac{1}{2}\nu^{K}{}_{L}\{\mathbf{A},\mathbf{V}^{L}{}_{K}
\}_{QPB}=\frac{1}{2i\hbar}\nu^{K}{}_{L}[\mathbf{A},\mathbf{V}^{L}{}_{K}], \label{q.108b}\\
\delta\mathbf{A}&=&\frac{1}{2}\omega^{a}{}_{b}\{\mathbf{A},{\boldsymbol\varrho}^{b}{}_{a}
\}_{QPB}=\frac{1}{2i\hbar}\omega^{a}{}_{b}[\mathbf{A},{\boldsymbol\varrho}^{b}{}_{a}], \label{q.108c}\\
\delta\mathbf{A}&=&\frac{1}{2}\omega^{a}{}_{b}\{\mathbf{A},{\boldsymbol\tau}^{b}{}_{a}
\}_{QPB}=\frac{1}{2i\hbar}\omega^{a}{}_{b}[\mathbf{A},{\boldsymbol\tau}^{b}{}_{a}]. \label{q.108d}
\end{eqnarray}
As usual, $[\mathbf{A},\mathbf{B}]$ denotes the commutator and $\{\mathbf{A},\mathbf{B}\}_{\rm QPB}$ is the corresponding quantum Poisson bracket, $[\mathbf{A},\mathbf{B}]=\mathbf{A}\mathbf{B}-\mathbf{B}\mathbf{A}$, $\{\mathbf{A},\mathbf{B}\}_{\rm QPB}=(1/i\hbar)[\mathbf{A},\mathbf{B}]$.
Again the $1/2$-factor follows from the use of summation convention over the whole range of indices $i$, $j$, $K$, $L$, $a$, $b$, in a consequence of the skew-symmetry of coefficients $\mu^{i}{}_{j}$, $\nu^{K}{}_{L}$, $\omega^{a}{}_{b}$. Let us remind that the formulas analogous to the left-hand sides of (\ref{q.108a})--(\ref{q.108d}) hold in classical theory. Formally one should then replace all operators by the corresponding phase space functions and the quantum Poisson bracket symbol by the classical one.

Just like in the classical theory, it is convenient to introduce a partial diagonalization of the doubly-invariant expression for the kinetic energy:
$\mathbf{M}^{a}{}_{b}=-{\boldsymbol\varrho}^{a}{}_{b}-
{\boldsymbol\tau}^{a}{}_{b}$, $\mathbf{N}^{a}{}_{b}={\boldsymbol\varrho}^{a}{}_{b}-
{\boldsymbol\tau}^{a}{}_{b}$. It must be stressed however that for the spatial dimensions higher than $2$, $n>2$, these quantities fail to be constants of motion even in the geodetic and doubly invariant situations. However, just like in the corresponding classical problems, the Casimir invariants built of ${\boldsymbol\varrho}^{a}{}_{b}$, ${\boldsymbol\tau}^{a}{}_{b}$ are constants of motion even in the non-geodetic models with potential energies depending only on deformation invariants.

Let us also stress that similarly to the classical model, the affinely invariant kinetic energies may be expressed by the Casimir invariants as follows:
\begin{eqnarray}
\mathbf{T}^{\rm aff-aff}_{\rm int}&=&\frac{1}{2A}\mathbf{C}(2)-
\frac{B}{2A(A+nB)}\mathbf{p}^{2}, \label{q.111}\\
\left\{\begin{array}{c}
\mathbf{T}^{\rm met-aff}_{\rm int}\\
\mathbf{T}^{\rm aff-met}_{\rm int}
\end{array}\right\}&=&\frac{1}{2\alpha}\mathbf{C}(2)+
\frac{1}{2\beta}\mathbf{p}^{2}+\frac{1}{2\mu}
\left\{\begin{array}{c}
\|\mathbf{S}\|^{2}\\
\|\mathbf{V}\|^{2}
\end{array}\right\}, \label{q.112-113}
\end{eqnarray}
where
\begin{equation}\label{q.114}
\|\mathbf{S}\|^{2}=-\frac{1}{2}\mathbf{S}^{a}{}_{b}\mathbf{S}^{b}{}_{a},\qquad
\|\mathbf{V}\|^{2}=-\frac{1}{2}\mathbf{V}^{A}{}_{B}\mathbf{V}^{B}{}_{A}.
\end{equation}
Just like in the corresponding classical formulas the inertial constants $\alpha$, $\beta$, $\mu$ are given by $\alpha=I+A$, $\beta=-(I+A)(I+A+nB)/B$, $\mu=(I^{2}-A^{2})/I$. Obviously, just like in the classical theory, $1/\beta=0$, $1/\mu=0$ if $B=0$; in this sense our notation is not very happy. $\mathbf{C}(2)$ is the second Casimir invariant of ${\rm GL}(n,\mathbb{R})$, and more generally we have $C(k)=\boldsymbol\Sigma^{a}{}_{b}\boldsymbol\Sigma^{b}{}_{c}\ldots\boldsymbol
\Sigma^{r}{}_{s}\boldsymbol\Sigma^{s}{}_{a}=
\widehat{\boldsymbol\Sigma}^{a}{}_{b}\widehat{\boldsymbol\Sigma}^{b}{}_{c}
\ldots\widehat{\boldsymbol\Sigma}
^{r}{}_{s}\widehat{\boldsymbol\Sigma}^{s}{}_{a}$, where each $C(k)$ contains the product of $k$ $\boldsymbol\Sigma$-s or $\widehat{\boldsymbol\Sigma}$-s.

In certain problems it is convenient to use the Casimir operators of ${\rm SL}(n,\mathbb{R})$ instead of ${\rm GL}(n,\mathbb{R})$. Then we obtain slightly modified formulas for the affinely-invariant kinetic energies:
\begin{eqnarray}
\mathbf{T}^{\rm aff-aff}_{\rm int}&=&\frac{1}{2A}\mathbf{C}_{{\rm SL}(n)}(2)+
\frac{1}{2n(A+nB)}{\mathbf{p}}^{2},\\
\left\{\begin{array}{c}
\mathbf{T}^{\rm met-aff}_{\rm int}\\
\mathbf{T}^{\rm aff-met}_{\rm int}
\end{array}\right\}&=&\frac{1}{2\alpha}\mathbf{C}_{{\rm SL}(n)}(2)+\frac{1}{2\widetilde{\beta}}\mathbf{p}^{2}+\frac{1}{2\mu}
\left\{\begin{array}{c}
\|\mathbf{S}\|^{2}\\
\|\mathbf{V}\|^{2}
\end{array}\right\},
\end{eqnarray}
where the modified inertial constant $\widetilde{\beta}$ is given by the following expression: $\widetilde{\beta}=n(I+A+nB)$. The Casimir invariant of ${\rm SL}(n,\mathbb{R})$ reads $\mathbf{C}_{{\rm SL}(n)}(k)={\boldsymbol\sigma}^{a}{}_{b}{\boldsymbol\sigma}^{b}{}_{c}\ldots {\boldsymbol\sigma}^{r}{}_{s}{\boldsymbol\sigma}^{s}{}_{a}=\widehat{{\boldsymbol\sigma}}^{a}{}_{b}
\widehat{{\boldsymbol\sigma}}^{b}{}_{c}\ldots\widehat{{\boldsymbol\sigma}}
^{r}{}_{s}\widehat{{\boldsymbol\sigma}}^{s}{}_{a}$, where the ${\boldsymbol\sigma}$- and $\widehat{{\boldsymbol\sigma}}$-operators are given by (\ref{q.97}). In the above formulas the isochoric (volume-preserving) parts and the pure dilatations are mutually separated from each other.

Let us mention that in spite of the non-compactness of the unimodular group there are discrete spectrum solutions for the purely geodetic models, without any extra introduced potential. There is even a characteristic threshold between the discrete and continuous spectrum. It is so as if the kinetic energy itself was used to model the potential interactions. This resembles the Maupertuis principle, however now there are no ``tricks'' with introducing potentials; they have their origin in symmetry principles. Only the purely dilatation potential must be stabilized, e.g., by some attractive term. But of course even more general doubly isotropic potentials of the form $V\left(q^{1},\ldots,q^{n}\right)$ are acceptable and admit some kind of separation of variables procedure, just like in the classical theory. Nevertheless, the existence of discrete purely geodetic spectrum is very interesting in itself when the configuration space is non-compact.

It is well known that functions on compact groups may be expanded with respect to the matrix elements of irreducible representations \cite{Wey-28,Wey-31}. Of course, neither ${\rm GL}(n,\mathbb{R})$ nor ${\rm SL}(n,\mathbb{R})$ are compact, but one can use the two-polar decomposition with its compact factors ${\rm SO}(n,\mathbb{R})$. This is especially efficient in the special physical case $n=3$, because one knows a lot about the compact geometry of ${\rm SO}(3,\mathbb{R})$. In the planar case $n=2$ it is even more apparent, although the commutative structure of ${\rm SO}(2,\mathbb{R})$ leads to certain drastic simplifications. Let us begin with the general value of $n$. Let the $N(\alpha)\times N(\alpha)$ and $N(\beta)\times N(\beta)$ denote the dimensions of quadratic matrices $D(\alpha)$, $D(\beta)$ representing the elements of $G$ within the $\alpha$-th and $\beta$-th classes of irreducible representations. Let $\Omega$ denote the set of all these classes; obviously we mean the classes of mutually non-equivalent representations. Then every function on ${\rm GL}(n,\mathbb{R})\simeq{\rm LI}(U,V)$ may be uniquely expanded as follows:
\begin{equation}\label{q.121}
\Psi(\varphi)=\Psi(L,D,R)=\sum\limits_{\alpha,\beta\in\Omega}\sum
\limits^{N(\alpha)}_{m,n=1}\sum\limits^{N(\beta)}_{k,l=1}D^{\alpha}_{mn}(L)
f^{\alpha\beta}_{nk|ml}(D)D^{\beta}_{kl}\left(R^{-1}\right).
\end{equation}
Obviously, the expansion coefficients $f^{\alpha\beta}_{nk|ml}$ are constants as functions of $L$, $R$ but evidently depend on deformation invariants $D$, or equivalently $Q^{a}$, $q^{a}$.

As mentioned previously, the two-polar decomposition is not unique and this fact must be taken into account in (\ref{q.121}). Unfortunately, $L$, $R$ are multidimensional rotation matrices, what complicates the description in a remarkable way. One must distinguish situations when there is no equality between any pair of $q^{1},\ldots,q^{n}$ and when at least one equality takes place. The first case is much more easy. It is clear that for any matrix $W\in{\rm SO}(n,\mathbb{R})$ which in every row and column has only one $\pm 1$ element and remaining ones do vanish, the following holds:
$LWDW^{-1}R^{-1}=LD_{\rm perm}R^{-1}$, where $D_{\rm perm}$ is diagonal and differs from $D$ by the permutation of diagonal elements. Therefore, $f^{\alpha\beta}_{nk|ml}$ depend on deformation invariants and we must have the expression
\begin{equation}\label{q.123}
f^{\alpha\beta}_{nk|ml}\left(q^{\pi_{W}(1},\ldots,q^{n)}\right)=
\sum\limits^{N(\alpha)}_{r=1}\sum\limits^{N(\beta)}_{s=1}D^{\alpha}_{nr}(W)
f^{\alpha\beta}_{rs|ml}\left(q^{1},\ldots,q^{n}\right)D^{\beta}_{sk}(W)
\end{equation}
for any matrix $W$ of the mentioned type. The same is true on the subsets $M^{\left(k;p_{1},\ldots,p_{k}\right)}\subset{\rm SO}(n,\mathbb{R})\times \mathbb{R}\times{\rm SO}(n,\mathbb{R})$, where $\left(q^{1},\ldots,q^{n}\right)$ is degenerate, i.e., there is some coincidence between $\left(q^{1},\ldots,q^{n}\right)$. Then $W$ contains some continuous part. The special and the simplest case is the total degeneracy of deformation invariants when only $LR^{-1}$ is well defined whereas $L$, $R$ separately are not determined. Obviously, then we have $D=\lambda I_{n}$, $\lambda>0$ and $I_{n}$ is the identity matrix $n\times n$.

It is clear that spin and vorticity Casimir operators are respectively given by
$\mathbf{C}_{{\rm SO}(V,g)}(p)=\mathbf{S}^{i}{}_{k}\mathbf{S}^{k}{}_{m}\ldots
\mathbf{S}^{r}{}_{z}\mathbf{S}^{z}{}_{i}$, $\mathbf{C}_{{\rm SO}(U,\eta)}(p)=\mathbf{V}^{A}{}_{K}\mathbf{V}^{K}{}_{M}\ldots
\mathbf{V}^{R}{}_{Z}\mathbf{V}^{Z}{}_{A}$, where in every expression the number of factors $p$ is even and not greater than $n$. To be more precise, those greater than $n$ would lead to operators algebraically built of those with $p\leq n$. In expressions (\ref{q.114}) we were dealing with operators proportional to those with $p=2$ and in the dimension $n=3$ this is the only possibility. 

When $\alpha$, $\beta$, $m$, $l$ are kept fixed, we can omit the symbols $m$, $l$ in (\ref{q.123}) and simply use the reduced matrix form:
\begin{equation}\label{q.125}
\Psi(\varphi)=\Psi^{\alpha\beta}_{ml}(L,D,R)=
\sum\limits^{N(\alpha)}_{n=1}\sum\limits^{N(\beta)}_{k=1}
D^{\alpha}_{mn}(L)f^{\alpha\beta}_{nk}(D)D^{\beta}_{kl}\left(R^{-1}\right),
\end{equation}
where $f^{\alpha\beta}_{nk}$ by its very nature is a matrix in the omitted pair of indices.

\section{Special case of three dimensions}

Let us specialize everything to the special case $n=3$. Then the skew-symmetric matrix of coefficients $\omega^{a}{}_{b}$ may be expressed in terms of the rotation vector $\overline{k}$, i.e., $\omega^{a}{}_{b}=-\varepsilon^{a}{}_{bc}k^{c}$, $k^{a}=-(1/2)\varepsilon^{a}{}_{b}{}^{c}\omega^{b}{}_{c}$, where the indices of the Ricci symbol are moved trivially with the help of the Kronecker metric $\delta_{ab}$. Obviously, the versor $\overline{n}=\overline{k}/k$ is the oriented axis of the rotation vector and the modulus $k$ is the rotation angle. There are formulas where from certain point of view, the doubled range of $k$, i.e., $[0,2\pi]$ instead of $[0,\pi]$ parametrizes the covering group ${\rm SU}(2)$.

Let us repeat that in the ${\rm SO}(3,\mathbb{R})$ case, the corresponding rotation matrix $W(\overline{k})$ is given by the mutually equivalent formulas: $W(\overline{k})=\exp\left(k^{a}E_{a}\right)=\sum^{\infty}_{m=0}
(1/m!)\left(k^{a}E_{a}\right)^{m}$, where $\left(E_{a}\right)^{b}{}_{c}=-\varepsilon_{a}{}^{b}{}_{c}$, or more explicitly:
$W(\overline{k})\overline{u}=\overline{u}+\overline{k}\times\overline{u}
+(1/2)\overline{k}\times\left(\overline{k}\times\overline{u}\right)
+\ldots$, $W(\overline{k})^{a}{}_{b}=\cos k\;\delta^{a}{}_{b}+
(1-\cos k)k^{a}k_{b}/k^{2}+\sin k\;\varepsilon^{a}{}_{bc}k^{c}/k$. The corresponding differential operators of the left and right regular translations on the group ${\rm SO}(3,\mathbb{R})$ have the following form:
\begin{equation}
\left\{\begin{array}{c}
\mathbf{L}_{a}\\
\mathbf{R}_{a}
\end{array}\right\}=\frac{k_{a}}{k}\frac{\partial}{\partial k}-\frac{1}{2}
{\rm ctg}\frac{k}{2}\varepsilon_{ab}{}^{c}k^{b}\mathbf{D}_{c}
\pm\frac{1}{2}\mathbf{D}_{a}, \label{q.129a-b}
\end{equation}
where $\mathbf{D}$ are operators of inner automorphisms:
$\mathbf{D}_{a}=\mathbf{L}_{a}-\mathbf{R}_{a}=
\varepsilon_{ab}{}^{c}k^{b}\partial/\partial k^{c}$. It is important that on ${\rm SO}(3,\mathbb{R})$ we have $W(\pi\overline{n})=W(-\pi\overline{n})=W(\pi\overline{n})^{-1}$
and for $k>\pi$ the values of $W(k\overline{n})$ are repeated. The manifold ${\rm SO}(3,\mathbb{R})$, just like any other ${\rm SO}(n,\mathbb{R})$ with $n\geq 3$, is doubly connected. For $n=3$ the covering manifold ${\rm Spin}(3)$ is isomorphic with the group ${\rm SU}(2)$. Parametrization with the help of rotation vector is still valid, however in ${\rm SU}(2)$ $k$ runs over the range $[0,2\pi]$, and we have $u(\overline{k})=\exp\left(k^{a}e_{a}\right)=
\cos(k/2)I_{2}-(k^{a}/k)\sin(k/2)i\sigma_{a}$, where $e_{a}=\sigma_{a}/2i$ and $\sigma_{a}$ are Pauli matrices:
\begin{equation}\label{q.134}
\sigma_{1}=\left[\begin{matrix}0&1\\1&0\end{matrix}\right],\qquad \sigma_{2}=\left[\begin{matrix}0&-i\\i&0\end{matrix}\right],\qquad \sigma_{3}=\left[\begin{matrix}1&0\\0&-1\end{matrix}\right].
\end{equation}
Now $u(\pi\overline{n})\neq u(-\pi\overline{n})$, but for any $\overline{n}$ we have $u(2\pi\overline{n})=-u(\overline{n})$
and the ${\rm SU}(2)$-manifold is simply-connected.

If $n=3$, we can use the following expressions for the Casimir invariants:
\begin{equation}\label{q.136}
\mathbf{C}_{{\rm SO}(V,g)}(2)=\mathbf{S}^{2}_{1}+\mathbf{S}^{2}_{2}
+\mathbf{S}^{2}_{3},\qquad \mathbf{C}_{{\rm SO}(U,\eta)}(2)=\mathbf{V}^{2}_{1}+\mathbf{V}^{2}_{2}
+\mathbf{V}^{2}_{3}.
\end{equation}
Roughly speaking, they are ${\rm SO}(3,\mathbb{R})$-Casimirs. They are well-known quantities and their family begins and terminates at $p=2$. On ${\rm SO}(3,\mathbb{R})$ $\Omega$ consists of non-negative integers and one uses the traditional symbols $s,j=0,1,2,\ldots$ for $\alpha$, $\beta$. Obviously, $N(s)=2s+1$, $N(j)=2j+1$. The indices $(m,n)$, $(k,l)$ are jumping by one from $-s$ to $s$ and from $-j$ to $j$ respectively. In ${\rm SU}(2)$ their range consists of non-negative half-integers and  integers, also jumping by one. 

Both in ${\rm SO}(3,\mathbb{R})$ and ${\rm SU}(2)$ the Haar measure is proportional to:
\begin{equation}\label{q.137}
d\mu(\overline{k})=\frac{4}{k^{2}}\sin^{2}\frac{k}{2}d_{3}\overline{k}=
4\sin^{2}\frac{k}{2}\sin\vartheta dk d\vartheta d\varphi,
\end{equation}
where $k$, $\vartheta$, $\varphi$ are polar coordinates on ${\rm SO}(3,\mathbb{R})/{\rm SU}(2)$. We use the same formula if the weight function is to equal one at the unit element, when $k=0$. However, if we want to normalize the total measure to the unity, then some constant factors appear.

The formulas (\ref{q.121}), (\ref{q.125}) become in the above scheme of coefficients as
\begin{eqnarray}
&&\Psi(\varphi)=\Psi(L,D,R)=\sum\limits^{\infty}_{s,j=0}\sum
\limits^{s}_{m,n=-s}\sum\limits^{j}_{k,l=-j}D^{s}_{mn}(L)
f^{sj}_{nk|ml}(D)D^{j}_{kl}\left(R^{-1}\right),\quad\  \label{q.138}\\
&&\Psi(\varphi)=\Psi^{sj}_{ml}(L,D,R)=\sum
\limits^{s}_{n=-s}\sum\limits^{j}_{k=-j}D^{s}_{mn}(L)
f^{sj}_{nk}(D)D^{j}_{kl}\left(R^{-1}\right). \label{q.139}
\end{eqnarray}
The reduced amplitudes $\Psi^{sj}_{ml}$ satisfy the eigenequations of rotational Casimir invariants: $\|\mathbf{S}^{2}\|\Psi^{sj}_{ml}=\hbar^{2}s(s+1)\Psi^{sj}_{ml}$, 
$\|\mathbf{V}^{2}\|\Psi^{sj}_{ml}=\hbar^{2}j(j+1)\Psi^{sj}_{ml}$, where as previously $\|\mathbf{S}^{2}\|=\mathbf{S}^{2}_{1}+\mathbf{S}^{2}_{2}+\mathbf{S}^{2}_{3}$,
$\|\mathbf{V}^{2}\|=\mathbf{V}^{2}_{1}+\mathbf{V}^{2}_{2}+\mathbf{V}^{2}_{3}$.

Traditionally, one uses such a basis that the third components of the operators $\mathbf{S}$, $\mathbf{V}$ have fixed eigenvalues, i.e.,
$\mathbf{S}_{3}\Psi^{sj}_{ml}=\hbar m\Psi^{sj}_{ml}$,
$\mathbf{V}_{3}\Psi^{sj}_{ml}=\hbar l\Psi^{sj}_{ml}$. Similarly, for the operators ${\boldsymbol\varrho}_{3}$, ${\boldsymbol\tau}_{3}$ the following eigenequations are satisfied:
${\boldsymbol\varrho}_{3}\Psi^{sj}_{ml|nk}=\hbar n\Psi^{sj}_{ml|nk}$,
${\boldsymbol\tau}_{3}\Psi^{sj}_{ml|nk}=\hbar k\Psi^{sj}_{ml|nk}$.
Obviously, ${\boldsymbol\varrho}_{3}$, ${\boldsymbol\tau}_{3}$ are related to ${\boldsymbol\varrho}^{a}{}_{b}$, ${\boldsymbol\tau}^{a}{}_{b}$ just like $\mathbf{S}_{3}$, $\mathbf{V}_{3}$ are related to $\mathbf{S}^{a}{}_{b}$, $\mathbf{V}^{a}{}_{b}$ when $n=3$. 

When dealing with the configuration space diffeomorphic with ${\rm GL}^{+}(3,\mathbb{R})$, then obviously, $s$, $j$ are non-negative integers and $m$, $l$, $n$, $k$ are jumping by one from $-s$, $-j$ to $s$, $j$. But something similar may be done when the covering group $\overline{{\rm GL}^{+}(3,\mathbb{R})}$ is used as the configuration space, i.e., when we admit half-integer values of the quantum numbers $s$, $j$. One begins with the manifold ${\rm SU}(2)\times\mathbb{R}^{3}\times{\rm SU}(2)$ and with wave functions on this manifold. But of course ${\rm SU}(2)\times\mathbb{R}^{3}\times{\rm SU}(2)$ is not diffeomorphic with the covering group $\overline{{\rm GL}^{+}(3,\mathbb{R})}$. So, let us write for the wave functions the Peter-Weyl expansion:
\begin{equation}\label{q.144}
\Psi(u,q,v)=\sum\limits^{\infty}_{s,j\in \mathbb{N}/2\cup\{0\}}
\sum\limits^{s}_{m,n=-s}\sum\limits^{j}_{k,l=-j}D^{s}_{mn}(u)
f^{sj}_{nk|ml}(q)D^{j}_{kl}\left(v^{-1}\right),
\end{equation}
where the $s,j$-summation is extended over non-negative integers and half-inte\-gers, and the jumps of $m$, $n$, $k$, $l$ equal one. Obviously $\mathbb{N}$ denotes the set of all naturals and $\mathbb{N}/2$ --- the set of all naturals and half-naturals. But, as said above, this is a general expansion of $\Psi$ on ${\rm SU}(2)\times\mathbb{R}^{3}\times{\rm SU}(2)$. To obtain the expansion on $\overline{{\rm GL}^{+}(3,\mathbb{R})}$ and ${\rm GL}^{+}(3,\mathbb{R})$ one must introduce certain restrictions for the coefficients $f$ and for the very summation procedure. Namely, (\ref{q.144}) represents a wave function on $\overline{{\rm GL}^{+}(3,\mathbb{R})}$ only when the summation is extended over such set of $(s,j)$ which have the same ``halfness'', i.e., when they are simultaneously integers or simultaneously half-integers. Because only then their moduli are one-valued from the point of view of the quotient manifold ${\rm GL}^{+}(3,\mathbb{R})$. Of course, everything is based here on the assumption that they should be so, what one must accept, although it is not completely self-evident. Besides, some assumptions following from the non-uniqueness of the two-polar decomposition must be satisfied; we mean ones mentioned after (\ref{q.123}). The wave functions are single-valued in ${\rm GL}^{+}(3,\mathbb{R})$ when both $s$ and $j$ are integers. So, the series $\sum_{s,j\in\mathbb{N}/2\cup 0:j-s\in\mathbb{Z}}$ and $\sum_{s,j\in \mathbb{N}\cup\left\{0\right\}}$ are well defined respectively on $\overline{{\rm GL}^{+}(n,\mathbb{R})}$ and ${\rm GL}^{+}(n,\mathbb{R})$. And the superselection rule is necessary, namely the second series can not be combined with $\sum_{s,j}$ when $s=m+1/2$, $j=n+1/2$ where $m$, $n$ are non-negative integers. Of course again under the assumption that the moduli wave functions are one-valued functions from the point of view  of ${\rm GL}^{+}(n,\mathbb{R})$, i.e., are projectable from ${\rm GL}^{+}(n,\mathbb{R})$ to their quotients. Of course, we still are thinking about the special case $n=3$.

\section{Euclidean and affine models of kinetic energy}

Let us quote the quantum formula for the internal kinetic energy simultaneously that is left- and right-affinely invariant, i.e.,
\begin{eqnarray}
{\mathbf T}^{\rm aff-aff}_{\rm int}&=&-\frac{\hbar^{2}}{2A}{\mathbf D}_{\lambda}+\frac{\hbar^{2}B}{2A(A+nB)}\frac{\partial^{2}}{\partial q^{2}}\nonumber\\
&+& \frac{1}{32A}\sum_{a,b}\frac{\left({\mathbf M}^{a}{}_{b}\right)^{2}}{{\rm sh}^{2}[(q^{a}-q^{b})/2]}- \frac{1}{32A}\sum_{a,b}\frac{\left({\mathbf N}^{a}{}_{b}\right)^{2}}{{\rm ch}^{2}[(q^{a}-q^{b})/2]},\quad\label{q.145}
\end{eqnarray}
where $\mathbf{M}^{a}{}_{b}=-{\boldsymbol\rho}^{a}{}_{b}-{\boldsymbol\tau}^{a}{}_{b}$,  $\mathbf{N}^{a}{}_{b}={\boldsymbol\rho}^{a}{}_{b}-{\boldsymbol\tau}^{a}{}_{b}$, 
and
\begin{equation}
{\mathbf D}_{\lambda}=\frac{1}{P_{\lambda}}\sum_{a}\frac{\partial}{\partial q^{a}}P_{\lambda}\frac{\partial}{\partial q^{a}}= \sum_{a}\frac{\partial^{2}}{\partial (q^{a})^{2}}+\sum_{a}\frac{\partial \ln P_{\lambda}}{\partial q^{a}}\frac{\partial}{\partial q^{a}}, \label{q.147}
\end{equation}
$P_{\lambda}$ is given by (\ref{q.83}). The expression (\ref{q.145}) is the usual Laplace-Beltrami operator, but it is seen that besides of the ``naively'' expected term $\sum_{a}\partial^{2}/\partial (q^{a})^{2}$ it contains an additional first-order differential term with respect to $\partial/\partial q^{a}$. In any case, it is formally self-adjoint. One can in a sense eliminate the first-order term by the following substituting: $\varphi = \sqrt{P_{\lambda}}\, \Psi$.
Then the action of ${\mathbf D}_{\lambda}$ on $\Psi$ may be replaced by the action of some other operator $\widetilde{\bf D}_{\lambda}$ on $\varphi$:
\begin{equation}
- \frac{\hbar^{2}}{2A} \widetilde{\bf D}= - \frac{\hbar^{2}}{2A}\sum_{a}\frac{\partial^{2}}{\partial (q^{a})^{2}}+\widetilde{\mathbf{V}},\ 
\widetilde{\mathbf V}=-\frac{\hbar^{2}}{2A}\frac{1}{P_{\lambda}{}^{2}}+ \frac{\hbar^{2}}{4A}\frac{1}{P_{\lambda}}\sum_{a}\left(\frac{\partial P_{\lambda}}{\partial q^{a}}\right)^{2}\label{q.149-150}
\end{equation}
($\widetilde{\mathbf V}$ is the completely algebraic operator). In this way the first-order differential operator is eliminated. Nevertheless, the difficulty still exists but it is moved from kinetic energy to potential term, generating the ``bad'' term $\widetilde{\mathbf V}$.

In analogy to the classical theory the formula (\ref{q.145}) is replaced by the following ones for the spatially metrical-materially affine and spatially affine-materially metrical models:
\begin{eqnarray}
\left\{\begin{array}{c}
{\mathbf T}^{\rm met-aff}_{\rm int}\\
{\mathbf T}^{\rm aff-met}_{\rm int}
\end{array}\right\}&=&
-\frac{\hbar^{2}}{2\alpha}{\mathbf D}_{\lambda}- \frac{\hbar^{2}}{2\beta}\frac{\partial^{2}}{\partial q^{2}} 
+\frac{1}{2\mu}\left\{\begin{array}{c}
\left\|{\mathbf S}\right\|^{2}\\
\left\|{\mathbf V}\right\|^{2}
\end{array}\right\}\nonumber\\
&+&\frac{1}{32\alpha}\sum_{a,b}\frac{\left({\mathbf M}^{a}{}_{b}\right)^{2}}{{\rm sh}^{2}[(q^{a}-q^{b})/2]}-
\frac{1}{32\alpha}\sum_{a,b}\frac{\left({\mathbf N}^{a}{}_{b}\right)^{2}}{{\rm ch}^{2}[(q^{a}-q^{b})/2]}.\qquad \label{q.151-152}
\end{eqnarray}
It is seen that (\ref{q.145}) and (\ref{q.151-152}) differ in a rather cosmetic way, namely by the Casimirs of ${\mathbf S}^{a}$, ${\mathbf V}^{b}$ (${\mathbf S}^{a}{}_{b}$, ${\mathbf V}^{a}{}_{b}$). Similarly, for the doubly isotropic d'Alembert model with the scalar moment of inertia $I$ we obtain:
\begin{equation}
{\mathbf T}^{\rm d'A.}_{\rm int}= -\frac{\hbar^{2}}{2I}{\mathbf D}_{l} + \frac{1}{8I}\sum_{a,b}\frac{\left({\mathbf M}^{a}{}_{b}\right)^{2}}{\left(Q^{a}-Q^{b}\right)^{2}}+\frac{1}{8I}\sum_{a,b}\frac{\left({\mathbf N}^{a}{}_{b}\right)^{2}}{\left(Q^{a}+Q^{b}\right)^{2}},\label{q.153}
\end{equation}
where ${\mathbf D}_{l}$ is given by
\begin{equation}
{\mathbf D}_{l}=\frac{1}{P_{l}}\sum_{a}\frac{\partial}{\partial Q^{a}}P_{l}\frac{\partial}{\partial Q^{a}}= \sum_{a}\frac{\partial^{2}}{\partial \left(Q^{a}\right)^{2}}+ \sum_{a}\frac{\partial \ln P_{l}}{\partial Q^{a}}\frac{\partial}{\partial Q^{a}},\label{q.154}
\end{equation}
$P_{l}$ given by (\ref{q.83}). Again it is possible to eliminate from ${\mathbf D}_{l}$ the first-order derivatives by the substitution
$\varphi = \sqrt{P_{l}}\Psi$, but this introduces a new difficulty connected with the new ``potential'':
\begin{equation}
\widetilde{\bf V}_{l}=-\frac{\hbar}{2I}\frac{1}{P_{l}^{2}}+
\frac{\hbar^{2}}{4I}\frac{1}{P_{l}}\sum_{a}
\left(\frac{\partial P_{l}}{\partial Q^{a}}\right)^{2}.\label{q.156}
\end{equation} 
Obviously, (\ref{q.153})--(\ref{q.156}) may be seeen as a rather strange form of expressing the usual Laplace operator in the $n^{2}$-dimensional Euclidean space ${\mathbb R}^{n^{2}}$:
\begin{equation}
{\mathbf T}^{\rm d'A.}= -\frac{\hbar^{2}}{2I}\Delta^{n^{2}}=-\frac{\hbar^{2}}{2I}\sum_{i,A}
\frac{\partial^{2}}{\partial \left(\varphi^{i}{}_{A}\right)^{2}}.\label{q.157}
\end{equation}
However, the geodetic model based on (\ref{q.153}), (\ref{q.157}) is non-physical from the point of view of condensed matter, because it predicts only straight-line infinite motions in ${\mathbb R}^{n^{2}}$. Therefore, the apparently natural Cartesian coordinates are completely useless and one must use curvilinear ones, e.g., polar and two-polar ones, and introduce the potential energy model, first of all isotropic one, i.e., ${\mathbf H}={\mathbf T}^{\rm d'A.}+V\left(Q^{1}, \ldots , Q^{n}\right)$.

We have mentioned that for the affinely invariant models one can, in principle, describe elastic vibrations and dissociation threshold without using the potential $V$, i.e., basing merely on the geodetic Hamiltonian (kinetic energy). Nevertheless, the potential terms, first of all doubly isotropic ones, are also admissible. Moreover, we have seen that to describe correctly elastic vibrations one must introduce at least some dilatations-stabilizing potential depending only on the trace/determinant $q$. This fixes our special attention on the following highly-symmetric affine Hamiltonians:
\begin{equation}
\left\{\begin{array}{c}
{\mathbf H}^{\rm aff-aff}\\
{\mathbf H}^{\rm met-aff}\\
{\mathbf H}^{\rm aff-met}
\end{array}\right\}=
\left\{\begin{array}{c}
{\mathbf T}^{\rm aff-aff}\\
{\mathbf T}^{\rm met-aff}\\
{\mathbf T}^{\rm aff-met}
\end{array}\right\}
+{\mathbf V}\left(q^{1}, \ldots , q^{n}\right).\label{q.159a-c}
\end{equation}
This is very special class of Hamiltonians, Nevertheless, there are rigorously solvable among them, or at least ones suited to approximate solvability. This is due to the double isotropy of potentials and to the affine/metrical symmetry of kinetic energy. The essential point is that for the models (\ref{q.159a-c}) it is possible to perform a partial separation of variables and reduction to some special functions on the rotation group. Just like in the classical theory this is based on the two-polar decomposition of variables. 

Let us denote the matrix generators of $D^{\alpha}$ by $M^{\alpha}$, so that for any matrix $W(\omega)=\exp\left(\omega^{a}{}_{b}E^{b}{}_{a}/2\right)$ the representing matrix $D^{\alpha}(\omega)$ is given by $D^{\alpha}(\omega)=\exp\left(\omega^{a}{}_{b}M^{\alpha b}{}_{a}/2\right)$.
In three dimensions, using the dual pseudovector $\omega^{a}$, we have
$D^{j}(\omega)=\exp \left(\omega^{a} M^{j}{}_{a}\right)$. Obviously,
$\left[M^{j}{}_{a}, M^{j}{}_{b}\right]=-\varepsilon_{ab}{}^{c}M^{j}{}_{c}$.
Then we introduce Hermitian matrices of the $j$-th angular momentum as follows:
$S^{\alpha a}{}_{b}=(\hbar/i)M^{\alpha a}{}_{b}$, $S^{j}{}_{a}=(\hbar/i)M^{j}{}_{a}$. In three dimensions the commutation rules have the well-known form: $(1/i\hbar)[S^{j}{}_{a}, S^{j}{}_{b}]=\varepsilon_{ab}{}^{c}S^{j}{}_{c}$.

The advantage of doubly isotropic models is that the action of differential operators like ${\boldsymbol\rho}^{a}{}_{b}$, ${\boldsymbol\tau}^{a}{}_{b}$, ${\mathbf M}^{a}{}_{b}$, ${\mathbf N}^{a}{}_{b}$ becomes algebraized, similarly like the usual differentiation in Fourier representation. Let us introduce the following symbols for the left- and right-hand-side effective action on the amplitudes $f^{\alpha \beta}$, i.e.,
$\overrightarrow{S^{\alpha}}^{ a}{}_{b}f^{\alpha \beta}:=S^{\alpha a}{}_{b}f^{\alpha \beta}$, $\overleftarrow{S^{\beta}}^{ a}{}_{b}f^{\alpha \beta}:=f^{\alpha \beta}S^{\beta a}{}_{b}$. It is clear that the action of differential operators $\left({\mathbf M}^{a}{}_{b}\right)^{2}$, $\left({\mathbf N}^{a}{}_{b}\right)^{2}$ is represented in the two-polar decomposition language respectively by the action of $(-\overleftarrow{S^{\beta}}^{a}{}_{b}-
\overrightarrow{S^{\alpha}}^{a}{}_{b})^{2}$ and $(\overleftarrow{S^{\beta}}^{a}{}_{b}-
\overrightarrow{S^{\alpha}}^{a}{}_{b})^{2}$ on the reduced amplitudes $f^{\alpha \beta}$.

Let us consider the stationary Schr\"{o}dinger equation ${\mathbf H}\Psi=E\Psi$ 
with the left affinely invariant and right-isotropic or the left-isotropic and right affinely invariant kinetic energy and with doubly isotropic potential energy. Obviously, it is not only admitted by just considered on the first place, the model of internal kinetic energy with the simultaneous  left and right full affine invariance. For simplicity we consider here only the internal part of the Schr\"{o}dinger equation. Under our invariance assumptions the above equation is equivalent to some infinite discrete family of eigenequations for the reduced amplitudes $H^{\alpha \beta} f^{\alpha \beta}=E^{\alpha \beta}f^{\alpha \beta}$, where for any $\alpha, \beta \in \Omega$, $f^{\alpha \beta}$ is again the $N(\alpha)\times N(\beta)$ matrix depending on deformation invariants $q^{a}$. The above eigenvalue problem is $N(\alpha)\times N(\beta)$-fold degenerate; let us remind that $N(\alpha)$ is the dimension of the $\alpha$-th representation. In the previously used symbols $f^{\alpha\beta}_{nk|ml}$ the indices $m,l$ have to do with the mentioned degeneracy of solutions for $f^{\alpha \beta}_{nk}$. The symbol ${\mathbf H}^{\alpha \beta}$ is an $N(\alpha)\times N(\beta)$ matrix the elements of which are differential operators ${\mathbf H}^{\alpha \beta}={\mathbf T}^{\alpha \beta}+{\mathbf V}$, 
where ${\mathbf V}$ denotes a doubly-isotropic potential energy or at least a dilatation-stabilizing potential. ${\mathbf T}^{\alpha \beta}$ is the kinetic energy operator restricted to the subspace labeled by the eigenvalues $(\alpha, \beta)$ of the general Hilbert space. We have assumed that the representations $D^{\alpha}$ of SO$(n, {\mathbb R})$ are irreducible. Let us construct the matrices $C^{\alpha}(p)$ given by $C^{\alpha}(p)^{a}{}_{z}:=S^{\alpha a}{}_{b}S^{\alpha b}{}_{c} \ldots S^{\alpha u}{}_{w}S^{\alpha w}{}_{z}$ ($p$ factors). Due to the irreducibility of $D^{\alpha}$ these matrices are proportional to the identity matrix, due to the Schur theorem: $C^{\alpha}(p)=(\hbar/i)^{p}C(\alpha, p){\rm I}_{N(\alpha)}$, where $C(\alpha, p)$ are eigenvalues of the Casimir operators built of the generators of the left and right regular translations in SO$(n, {\mathbb R})$.
After some calculations one can find that the reduced operators ${\mathbf T}^{\alpha \beta}$ for the affine-affine, metric-affine and the affine-metric operators of the kinetic energy are given by the following counterparts of expressions (\ref{q.145}) and (\ref{q.151-152}) respectively:
\begin{eqnarray}
&&{\mathbf T}^{\alpha \beta}_{\rm aff-aff}f^{\alpha \beta}=-\frac{\hbar^{2}}{2A}{\mathbf D}_{\lambda}f^{\alpha \beta}
+\frac{\hbar^{2}B}{2A(A+nB)}\frac{\partial^{2}}{\partial q^{2}}f^{\alpha \beta}\nonumber\\
&&\quad + \frac{1}{32A}\sum_{a,b} \frac{\left(\overleftarrow{S^{\beta}}^{a}{}_{b}-
\overrightarrow{S^{\alpha}}^{a}{}_{b}\right)^{2}}
{{\rm sh}^{2}[(q^{a}-q^{b})/2]}f^{\alpha \beta}-
\frac{1}{32A}\sum_{a,b} \frac{\left(\overleftarrow{S^{\beta}}^{a}{}_{b}+
\overrightarrow{S^{\alpha}}^{a}{}_{b}\right)^{2}}{{\rm ch}^{2}[(q^{a}-q^{b})/2]}f^{\alpha \beta},\label{q.171}\\
&&\left\{\begin{array}{c}
{\mathbf T}^{\alpha \beta}_{\rm met-aff}\\
{\mathbf T}^{\alpha \beta}_{\rm aff-met}
\end{array}\right\}f^{\alpha \beta}
=-\frac{\hbar^{2}}{2 \alpha}{\mathbf D}_{\lambda}f^{\alpha \beta}
-\frac{\hbar^{2}}{2\beta}\frac{\partial^{2}}{\partial q^{2}}f^{\alpha \beta}
-\frac{\hbar^{2}}{2\mu}
\left\{\begin{array}{c}
C(\alpha,2)\\
C(\beta,2)
\end{array}\right\}
f^{\alpha \beta}\nonumber\\
&&\quad + \frac{1}{32 \alpha}\sum_{a,b} \frac{\left(\overleftarrow{S^{\beta}}^{a}{}_{b}-
\overrightarrow{S^{\alpha}}^{a}{}_{b}\right)^{2}}
{{\rm sh}^{2}[(q^{a}-q^{b})/2]}f^{\alpha \beta}-\frac{1}{32 \alpha}\sum_{a,b} \frac{\left(\overleftarrow{S^{\beta}}^{a}{}_{b}+
\overrightarrow{S^{\alpha}}^{a}{}_{b}\right)^{2}}
{{\rm ch}^{2}[(q^{a}-q^{b})/2]}f^{\alpha \beta}.\label{q.172-173}
\end{eqnarray}
One must not confuse the representation labels $\alpha$, $\beta$ with the inverses of the multiplicative constants. We apologise for this inconvenience. It is seen that there is no very essential difference between those three expressions; only one in multiplicative constants and with the use of spin and vorticity Casimirs. Those formulas are valid for any spatial dimension $n$. In the directly physical case $n=3$ we have obviously $\alpha=s=0,1/2, 1, \ldots \in \mathbb{N}/2 \cup \left\{0\right\}$, $\beta=j=0,1/2, 1, \ldots \in \mathbb{N}/2 \cup \left\{0\right\}$ when we admit half-integer values of angular momenta and vorticity. If we admit only integer values, then obviously $s,j \in \mathbb{N} \cup \left\{0\right\}$. Obviously, in three dimensions we have $C(2,2)=s(s+1)$, $C(j)=j(j+1)$. Then the constant terms in the formulas (\ref{q.172-173}) are simply given by $\hbar^{2}s(s+1)/2\mu$, $\hbar^{2}j(j+1)/2\mu$. Those corrections to the affine-affine model are very interesting and have the structure interesting for any physicist. The term $\hbar^{2}s(s+1)/2\mu$ is interesting as the rotational connection to the situation when the purely deformative part is established and later on excited to quicker rotations. From this point of view the correction term $\hbar^{2}j(j+1)/2\mu$ in (\ref{q.172-173}) is perhaps even more interesting because it may be interpreted as a kind of internal quantum term following from the SO$(3, {\mathbb R})$-group or its covering SU$(2)$. This might be something like the isospin. To combine them, i.e., to obtain some combination of terms $\hbar^{2}s(s+1)/2\mu$, $\hbar^{2}j(j+1)/2\mu$, we should modify more deeply the primary affine-affine model.

Let us observe that the use of the two-polar description together with the Weyl-Peter theorem enables one to simplify the expression for the scalar product, reducing it to the integration over the $q^{i}$-variables and the series summation over discrete variables. Namely, if we take two wave functions $\Psi_{1}$, $\Psi_{2}$ with the deformation profiles $f_{1}$, $f_{2}$, then one can easily show that
\begin{equation}
\left\langle \Psi_{1}\left|\right.\Psi_{2}\right\rangle=\sum_{\alpha, \beta \in \Omega}\frac{1}{N(\alpha)N(\beta)}\int \sum_{n, m =1}^{N(\alpha)}\sum_{k, l =1}^{N(\beta)} {\overline{f_{1}}}^{\alpha\beta}_{nk|ml}
{f_{2}}^{\alpha\beta}_{nk|ml}P_{\lambda}dq^{1}\ldots dq^{n}. \label{q.174}
\end{equation}
When we restrict ourselves to the subspace of wave functions with fixed labels $\alpha, \beta, m, l$ and use the simplified $N(\alpha)\times N(\beta)$-matrix amplitudes of the form $\Psi^{\alpha \beta}\left(L; q^{1},\ldots, q^{n}; R\right)=D^{\alpha}(l) f^{\alpha \beta}\left(q^{1},\ldots, q^{n}\right)D^{\beta}(R^{-1})$, this scalar product may be reduced to the following expression:
\begin{equation}
\left\langle \Psi_{1}{}^{\alpha \beta}\left|\right.\Psi_{2}{}^{\alpha \beta}\right\rangle=\frac{1}{N(\alpha)N(\beta)}\int {\rm Tr}
\left(f_{1}^{\alpha \beta +}(q) f_{2}^{\alpha\beta}(q)\right)
P_{\lambda}(q)dq^{1}\ldots dq^{n}. \label{q.176}
\end{equation}
Similarly, for the general case it may be written as follows:
\begin{equation}
\left\langle \Psi_{1}{}\left|\right.\Psi_{2}{}\right\rangle=\sum_{\alpha, \beta \in \Omega}\frac{1}{N(\alpha)N(\beta)}\int {\rm Tr}\left(f_{1}^{\alpha \beta +} f_{2}^{\alpha \beta}\right)P_{\lambda} dq^{1}\ldots dq^{n}, \label{q.177}
\end{equation}
where, obviously, ${\rm Tr}\left(f_{1}^{\alpha \beta +} f_{2}^{\alpha \beta}\right)=\sum_{n, m =1}^{N(\alpha)}\sum_{k, l =1}^{N(\beta)} {\overline{f_{1}}}^{\alpha\beta}_{nk|ml}{f_{2}}^{\alpha\beta}_{nk|ml}$;
the weight factor $P_{\lambda}$ can be eliminated from (\ref{q.177}) by the substitution $\varphi = \sqrt{P_{l}}\Psi$.

Let us mention again the usual d'Alembert models. Now for the isotropic inertial tensor and for the doubly isotropic potential energy we can also write that the Schr\"{o}dinger equation ${\mathbf H}\Psi=E\Psi$ 
for the isotropic potentials reduces to the family ${\mathbf H}^{\alpha \beta}f^{\alpha \beta}=E^{\alpha \beta}f^{\alpha \beta}$, where
\begin{eqnarray}
{\mathbf H}^{\alpha \beta}_{\rm d'A.}f^{\alpha \beta}&=&-\frac{\hbar^{2}}{2I}{\mathbf D}_{l}f^{\alpha \beta}+ \frac{1}{8I}\sum_{a,b} \frac{\left(\overleftarrow{S^{\beta}}^{a}{}_{b}-
\overrightarrow{S^{\alpha}}^{a}{}_{b}\right)}
{\left(Q^{a}-Q^{b}\right)^{2}}f^{\alpha \beta}\nonumber\\
&+&\frac{1}{8I}\sum_{a,b} \frac{\left(\overleftarrow{S^{\beta}}^{a}{}_{b}+
\overrightarrow{S^{\alpha}}^{a}{}_{b}\right)^{2}}
{\left(Q^{a}+Q^{b}\right)^{2}}f^{\alpha \beta}+V\left(Q^{1}, \ldots ,Q^{n}\right)f^{\alpha \beta}.\label{q.181}
\end{eqnarray}
It is clear that without the potential term, i.e., when dealing with the geodetic model, all motions are infinite and there are no elastic vibrations, just like in the corresponding classical theory.

\section{Special case of two dimensions}

We have seen that in classical mechanics the geodetic affinely-invariant models on ${\rm SL}(n, {\mathbb R})$ may describe elastic vibrations. Moreover, there exists a sharp threshold between finite vibrations and infinite escaping motions. It is given by some relationship between spin and vorticity. In ${\rm GL}(n, {\mathbb R})$ the same qualitative picture may be obtained by introducing some stabilizing dilatational potential. By analogy something similar exists in quantum theory. Let us consider this again in the special, particularly simple model in $n=2$. 

The Haar measure on ${\rm GL}(2, {\mathbb R})$ may be expressed as $d\lambda\left(\alpha; q^{1}, q^{2}; \beta \right)= \left|{\rm sh} \left(q^{1} - q^{2}\right)\right|d\alpha \, d\beta \, dq^{1} dq^{2}$, 
where, as usual $q^{1}$, $q^{2}$ are logarithmic deformation invariants and $\alpha$, $\beta$ are polar angles parametrizing respectively $L$ and $R$ in the two-polar decomposition. As usual we introduce new variables:
$q=\left(q^{1} + q^{2}\right)/2$, $x= q^{2} - q^{1}$.
In certain problems it is also convenient to introduce the following mixed angular variables: $\gamma=\left(\beta - \alpha \right)/2$, $\delta=\left(\beta + \alpha \right)/2$. Therefore, $d\lambda\left(\alpha; q, x; \beta \right)= \left|{\rm sh}\, x\right|d\alpha \, d\beta \, dq dx$, $P_{\lambda}=\left|{\rm sh} \, x\right|$. 

According to the Peter-Weyl theorem, or more directly, to the Fourier theorem, we have the following expansion for our wave functions on ${\rm GL}(2, {\mathbb R})$:
\begin{equation}
\Psi \left(\alpha; q, x; \beta \right)= \sum_{m,n\in {\mathbb Z}}f^{mn}(q,x)e^{im\alpha}e^{in\beta}.\label{q.186}
\end{equation}
For the affine-affine, metric-affine, and affine-metric models $T^{\rm aff-aff}_{\rm int}$, $T^{\rm met-aff}_{\rm int}$, $T^{\rm aff-met}_{\rm int}$ we have the following reduced expressions for the kinetic energy ${\mathbf T}^{mn}$:
\begin{eqnarray}
{\mathbf T}^{mn}_{\rm aff-aff}f^{mn}&=&-\frac{\hbar^{2}}{A}{\mathbf D}_{x}f^{mn}- \frac{\hbar^{2}}{4(A+2B)}\frac{\partial^{2}f^{mn}}{\partial q^{2}}\nonumber\\
&+&\frac{\hbar^{2}(n-m)^{2}}{16A{\rm sh}^{2}(x/2)}f^{mn}-
\frac{\hbar^{2}(n+m)^{2}}{16A{\rm ch}^{2}(x/2)}f^{mn},\label{q.187}\\
\left\{\begin{array}{c}
{\mathbf T}^{mn}_{\rm met-aff}\\
{\mathbf T}^{mn}_{\rm aff-met}
\end{array}\right\}f^{mn}
&=&-\frac{\hbar^{2}}{\alpha}{\mathbf D}_{x}f^{mn}- \frac{\hbar^{2}}{2\widetilde{\beta}}\frac{\partial^{2}f^{mn}}{\partial q^{2}}+
\frac{\hbar^{2}}{\mu}
\left\{\begin{array}{c}
m^{2}\\
n^{2}
\end{array}\right\}f^{mn}\nonumber\\
&+&\frac{\hbar^{2}(n-m)^{2}}{16\alpha{\rm sh}^{2}(x/2)}f^{mn}-
\frac{\hbar^{2}(n+m)^{2}}{16\alpha{\rm ch}^{2}(x/2)}f^{mn},\label{q.188-189}
\end{eqnarray}
where
\begin{equation}
{\mathbf D}_{x}f^{mn}=\frac{1}{\left|{\rm sh} \, x\right|}\frac{\partial}{\partial x}\left(\left|{\rm sh} \, x\right|\frac{\partial f^{mn}}{\partial x}\right). \label{q.190}
\end{equation}

Of course, for the purely geodetic models on ${\rm GL}(2, {\mathbb R})$ the spectrum is continuous, because dilatational motion is free. To avoid this fact we must introduce to the Hamiltonian some dilatation-stabilizing potential $V_{\rm dil}(q)$. This may be either the potential well or some harmonic oscillator with large elastic constant. Obviously, the problem is also explicitly separable for any potential of the form:
$V(q,x)=V_{\rm dil}(q)+V_{\rm sh}(x)$.
The corresponding solutions of the time-independent Schr\"{o}dinger equation will be sought in the product form: $f^{mn}(q,x)=\varphi^{mn}(q)\chi^{mn}(x)$. 

It is interesting that there exists a discrete spectrum for $\chi$-terms in ${\rm SL}(2, {\mathbb R})$ even in the purely geodetic models without any shear potential $V_{\rm sh}(x)$. This depends on the mutual relationship between ``gyroscopic'' quantum numbers $m$, $n$. If $\left|n-m\right|<\left|n+m\right|$, then the attractive ${\rm ch}^{-2}$-term becomes dominant at large distances, when $\left|x\right| \rightarrow \infty$, and the spectrum for $\chi$ is then discrete. 

Conversely, it becomes continuous when $\left|n-m\right|>\left|n+m\right|$. For the affine-affine model (\ref{q.187}) the spectrum is not bounded from below. Conversely, for the affine-metric nad metric-affine models the kinetic energy may be bounded from below and so is the spectrum. This happens for certain open range of parameters $I$, $A$, and $B$.

Similar phenomena hold for the dimension of space greater than $2$, because everything follows from the commutation rules (structure constants) of ${\rm SL}(n, {\mathbb R})$.

\end{document}